\documentclass[11pt]{article}
\usepackage[usenames,dvipsnames,svgnames,table]{xcolor} 
\usepackage[obeyspaces,hyphens,spaces]{url}
\usepackage{jcapmod}
\usepackage{geometry}
\usepackage{changepage}
\newcommand{\be}{\begin{equation}}
\newcommand{\ee}{\end{equation}}

\usepackage{tabu}
\usepackage{booktabs}
\usepackage[english]{babel}
\usepackage{amsmath, amssymb, amsbsy, amstext, amsthm, simplewick}
\usepackage{hyperref}
\usepackage{graphicx}
\usepackage{amsfonts}
\usepackage{enumitem}
\usepackage{upgreek}
\usepackage{framed}
\usepackage{tensor}
\usepackage{pifont}
\usepackage{latexsym, mathrsfs}
\usepackage{array}
\usepackage{hyperref}
\usepackage{xspace}
\usepackage{longtable}
\usepackage{multirow}
\usepackage{cases}
\usepackage{empheq}
\usepackage{bm}
\usepackage{bbm}
\usepackage[table]{xcolor}
\usepackage[dvipsnames]{xcolor}
\usepackage{colortbl}
\usepackage{tablefootnote}
\usepackage{braket}
\usepackage[utf8]{inputenc}
\allowdisplaybreaks
\usepackage{kantlipsum}
\usepackage{graphicx}
\usepackage{subcaption}
\usepackage{adjustbox}

\usepackage[load-configurations=astronomy, range-units=brackets, range-phrase=-, per-mode=reciprocal, mode=math]{siunitx}
\usepackage{threeparttable}

\usepackage{ragged2e}

\usepackage[normalem]{ulem}

\usepackage{hhline}
\usepackage{array}
\newcolumntype{P}[1]{>{\centering\arraybackslash}p{#1}}
\usepackage{booktabs}
\usepackage{tikz}
\usetikzlibrary{calc,shadings,patterns,tikzmark,fadings}

\definecolor{Blue}{rgb}{0.25, 0.41, 0.88}
\definecolor{Red}{rgb}{0.92,0.,0.}
\definecolor{darkorange}{rgb}{1.0,0.549,0.}
\definecolor{cobalt}{RGB}{44, 98, 120}
\definecolor{Mathematica1}{rgb}{0.368417, 0.506779, 0.709798}
\definecolor{Mathematica2}{rgb}{0.880722, 0.611041, 0.142051}
\definecolor{Mathematica3}{rgb}{0.560181, 0.691569, 0.194885}
\definecolor{Mathematica4}{rgb}{0.922526, 0.385626, 0.209179}
\definecolor{Mathematica5}{rgb}{0.528488, 0.470624, 0.701351}
\definecolor{Mathematica6}{rgb}{0.772079, 0.431554, 0.102387}
\definecolor{Mathematica7}{rgb}{0.363898, 0.618501, 0.782349}
\definecolor{Mathematica8}{rgb}{1, 0.75, 0}
\definecolor{Mathematica9}{rgb}{0.647624, 0.37816, 0.614037}
\definecolor{plotBlue}{RGB}{94, 130, 181}
\definecolor{plotRed}{RGB}{233, 85, 54}
\definecolor{plotGreen}{RGB}{142, 176, 50}
\definecolor{plotPurple}{RGB}{135, 120, 178}

\renewcommand\_{\textunderscore\allowbreak}

\definecolor{cornellRed}{HTML}{B31B1B}
\definecolor{cornellBlue}{HTML}{0068AC}
\definecolor{cornellGreen}{HTML}{6EB43F}
\definecolor{purple}{HTML}{66023C}

\newcolumntype{C}[1]{>{\centering\let\newline\\\arraybackslash\hspace{0pt}}m{#1}}

\def\d{{\rm d}}

\newcommand{\affsize}{10}

\makeatletter
\newlength{\apb@width}
\newcommand{\autoparbox}[2][c]{\settowidth{\apb@width}{#2}\parbox[#1]{\apb@width}{#2}}

\makeatother

\numberwithin{equation}{section}

\def\beq{\begin{equation}}
\def\eeq{\end{equation}}

\def\bea{\begin{eqnarray}}
\def\eea{\end{eqnarray}}

\def\d{{\rm d}}

\def\beq{\begin{equation}}
\def\eeq{\end{equation}}
\def\bea{\begin{eqnarray}}
\def\eea{\end{eqnarray}}

\def\d{{\rm d}}

\def\d{{\rm d}}

\DeclareRobustCommand{\SkipTocEntry}[4]{}

\usepackage{colortbl}
\definecolor{blue2}{cmyk}{1, 0.1, 0.1, 0}

\definecolor{pyBlue}{RGB}{31, 119, 180}
\definecolor{pyRed}{RGB}{214, 39, 40}
\definecolor{pyGreen}{RGB}{44, 160, 44}
\definecolor{pyBlue2}{RGB}{0, 111, 237}
\definecolor{pyRed2}{RGB}{224, 52, 36}

\usepackage{colortbl}
\definecolor{summersky}{cmyk}{0.71,0.33,0,0.5}
\definecolor{flamingo}{cmyk}{0,0.51,0.71,0.5}
\definecolor{rp}{cmyk}{0.2, 1, 0.6, 0}
\definecolor{pacificblue}{cmyk}{0.95,0.3,0, 0.5}
\definecolor{gray60}{cmyk}{0.4,0.4,0,0.8}

\renewcommand{\(}{\left(}
\renewcommand{\)}{\right)}
\renewcommand{\[}{\left[}
\renewcommand{\]}{\right]}

\usepackage[framemethod=default]{mdframed}
\newmdenv[skipabove=7pt,
skipbelow=7pt,
rightline=false,
leftline=false,
topline=false,
bottomline=false,
backgroundcolor=pacificblue!8,
linecolor=gray,
innerleftmargin=5pt,
innerrightmargin=5pt,
innertopmargin=10pt,
innerbottommargin=10pt,
leftmargin=0cm,
rightmargin=0cm,
linewidth=4pt]{eBox}

\newmdenv[skipabove=7pt,
skipbelow=7pt,
rightline=false,
leftline=false,
topline=false,
bottomline=false,
backgroundcolor=gray!10,
linecolor=gray,
innerleftmargin=5pt,
innerrightmargin=5pt,
innertopmargin=-5pt,
innerbottommargin=5pt,
leftmargin=0cm,
rightmargin=0cm,
linewidth=4pt]{eBox2}

\makeatletter
\def\Ddots{\mathinner{\mkern1mu\raise\p@
\vbox{\kern7\p@\hbox{.}}\mkern2mu
\raise4\p@\hbox{.}\mkern2mu\raise7\p@\hbox{.}\mkern1mu}}
\makeatother

\begin{document}

\pagenumbering{roman}

\begin{titlepage}

\newgeometry{vmargin={15mm}, hmargin={16mm,16mm}}

\baselineskip=15.5pt \thispagestyle{empty}
\begin{flushright}

\end{flushright}
\vspace{-0.5cm}

\begin{center}
\begin{minipage}{1.0\textwidth}
\centering

\begin{flushright}
MIT-CTP/5710
\end{flushright}

\vspace{0.5cm}

{\fontsize{15.5}{0}\selectfont \bfseries
Bring the Heat: Tidal Heating Constraints for Black Holes
\\[4pt]
and Exotic Compact Objects 
from the LIGO-Virgo-KAGRA Data 
}
 
\end{minipage}
\end{center}


\vspace{0.001cm}

\begin{center}
\begin{minipage}{1\textwidth}
\centering
{\fontsize{13}{0} \selectfont Horng Sheng Chia$^{1a}$, Zihan Zhou$^{2b}$, and Mikhail M. Ivanov$^{3c}$} 
\end{minipage}
\end{center}


\begin{center}
\begin{minipage}[c]{1.0\textwidth}
\centering

\textsl{\fontsize{\affsize}{0}\selectfont $^1$ School of Natural Sciences, Institute for Advanced Study, Princeton, NJ 08540, USA}

\textsl{\fontsize{\affsize}{0}\selectfont $^2$ Department of Physics, Princeton University, Princeton, NJ 08540, USA}

\textsl{\fontsize{\affsize}{0} \selectfont $^3$ Center for Theoretical Physics, Massachusetts Institute of Technology, Cambridge, MA 02139, USA}

\end{minipage}
\end{center}

\vspace{0.3cm}

\begin{center}
\begin{minipage}{0.88\textwidth}
\hrule \vspace{10pt}
\noindent {\bf Abstract} \\[0.02cm]

We present the first constraints on tidal heating for the binary systems detected in the LIGO-Virgo-KAGRA (LVK) gravitational wave data. Tidal heating, also known as tidal dissipation, characterizes the viscous nature of an astrophysical body and provides a channel for exchanging energy and angular momentum with the tidal environment. Using the worldline effective field theory formalism, we introduce a physically motivated and easily interpretable parametrization of tidal heating valid for an arbitrary compact  astrophysical object. We then derive the imprints of the spin-independent and linear-in-spin tidal heating effects of generic binary components on the waveform phases and amplitudes of quasi-circular orbits. Notably, the mass-weighted spin-independent tidal heating coefficient derived in this work, $\mathcal{H}_0$, is the dissipative analog of the tidal Love number. We constrain the tidal heating coefficients using the public LVK O1-O3 data. Our parameter estimation study includes two separate analyses: the first treats the catalog of binary events as binary black holes (BBH), while the second makes no assumption about the nature of the binary constituents and can therefore be interpreted as constraints for exotic compact objects. In the former case, we combine the posterior distributions of the individual BBH events and obtain a joint constraint of $-13 < \mathcal{H}_0 < 20$ at the $90\%$ credible interval for the BBH population. This translates into a bound on the fraction of the emitted gravitational wave energy lost due to tidal heating (or gained due to radiation enhancement effects) at
$|\Delta E_H/\Delta E_{\infty}|\lesssim 3\cdot 10^{-3}$. Our work provides the first robust framework for deriving and measuring tidal heating effects in merging binary systems, demonstrating its potential as a powerful probe of the nature of binary constituents and tests of new physics.

\vskip15pt
\hrule
\vskip10pt
\end{minipage}
\end{center}

\begin{center}
\begin{minipage}[l]{0.88\textwidth}
Email: $^{a}$\href{mailto:}{\texttt{hschia@ias.edu}}, $^{b}$\href{mailto:}{\texttt{zihanz@princeton.edu}}, $^{c}$\href{mailto:}{\texttt{ivanov99@mit.edu}}
\end{minipage}
\end{center}

\restoregeometry

\end{titlepage}

\thispagestyle{empty}
\setcounter{page}{2}
\tableofcontents

\pagenumbering{arabic}
\setcounter{page}{1}


\vspace{0.5cm}

\section{Introduction}

Gravitational wave (GW) science is a precision science. In the current growing era of GW astronomy, in which the LIGO-Virgo-KAGRA (LVK) detector network routinely detects orbiting binary black holes~\cite{LIGOScientific:2018mvr, LIGOScientific:2020ibl, LIGOScientific:2021usb, LIGOScientific:2021djp, Venumadhav:2019tad, Venumadhav:2019lyq, Olsen:2022pin, Nitz:2018imz, Nitz:2019hdf, Nitz:2021uxj, Nitz:2021zwj, Chia:2023tle, Mehta:2023zlk, Wadekar:2023gea},
the need for highly accurate and precise waveforms is increasingly apparent in order to achieve a wide range of GW science. 
To this end, one active area of research involves computing ever higher post-Newtonian (PN) terms for the waveform observables of merging binary systems \cite{Blanchet:2013haa,Blumlein:2021txe,Blumlein:2021txj,Cho:2022syn,Blanchet:2023soy,Trestini:2023ssa}. In this approach, a binary system is often first modeled as orbiting point particles, with their intrinsic parameters described by the component masses and spins. Moving beyond the point-particle approximation, various so-called finite-size effects which characterize the underlying multipolar structure of the binary constituents are incorporated in the waveform models. These finite-size effects include the spin-induced moments~\cite{Poisson:1997ha,Porto:2005ac,Marsat:2014xea, Levi:2014gsa,Levi:2015msa, Krishnendu:2017shb, Krishnendu:2018nqa,Chia:2020psj, Chia:2022rwc, Lyu:2023zxv}, the tidal deformability~\cite{love, Goldberger:2004jt, Flanagan:2007ix, Li:2007qu, Damour:2009vw, Binnington:2009bb, Vines:2011ud, Cardoso:2017cfl} and the tidal heating~\cite{hartle_heating, Poisson:1994yf, Tagoshi:1997jy, Alvi:2001mx, Hughes:2001jr, Poisson:2009di, Zahn:2008fk, Ogilvie:2014dwa, poisson_will_2014, murray1999solar} of the astrophysical bodies --- all of which would impact GW observables in non-trivial ways. In this paper, we conduct a comprehensive study on the imprints of tidal heating of general astrophysical bodies on GW observables. Our study \textit{a priori} makes no assumption about the nature of the compact objects, and is therefore applicable to both  binary black holes (BBH) and other general exotic types of compact binaries.


\vskip 4pt

Tidal heating~\cite{hartle_heating, Poisson:1994yf, Tagoshi:1997jy, Alvi:2001mx, Hughes:2001jr, Poisson:2009di, Zahn:2008fk, Ogilvie:2014dwa, poisson_will_2014, murray1999solar}, also commonly referred to as tidal dissipation, captures 
the viscosity of an astrophysical body and serves as an important channel for exchanging energy and angular momentum between the body and its external tidal environment. Perhaps the most familiar example of this effect lies literally at our cosmic backyard: the Earth-Moon system. In this case, the misalignment between the tidal bulge of the oceans on Earth and the radial separation between the Earth-Moon center of masses generates frictional forces that transfer energy and angular momentum between the two bodies and the orbit~\cite{poisson_will_2014, murray1999solar, Endlich:2015mke}. This process ultimately leads to tidal locking, whereby the rotational frequency of the Moon synchronizes with its orbital frequency around Earth, and is the reason why we only see one side of the Moon~\cite{poisson_will_2014, murray1999solar,Endlich:2015mke}. In addition to this well-known example, tidal dissipation is also responsible for a plethora of interesting astrophysical phenomena. For black holes orbiting around large companion stars, tidal heating has been invoked as a potential mechanism for spinning up the black holes~\cite{zahn1977, 2009A&A...497..243D, Zaldarriaga:2017qkw, Olejak:2021iux, Ma:2023nrf}, ultimately forming merging BBHs with at least one highly-spinning component which may be observable by the LVK detectors~\cite{Zackay:2019tzo, LIGOScientific:2020kqk, Chia:2021mxq, Roulet:2021hcu, KAGRA:2021duu}. Tidal dissipation is also directly responsible for a remarkable phenomenon known as black hole superradiance~\cite{zel1971generation, Zeldovich:1972spj, Starobinskil:1974nkd, Starobinsky:1973aij, Detweiler:1980uk, Bekenstein:1998nt, Dolan:2007mj, Brito:2015oca}, in which energy and angular momentum may be extracted from a rotating black hole if its spin is sufficiently high compared to the angular phase velocity of its surrounding perturbation field. Black hole superradiance has therefore been proposed as a powerful probe of ultralight dark matter~\cite{Arvanitaki:2009fg, Arvanitaki:2010sy, Witek:2012tr, Brito:2014wla, East:2017ovw, Baumann:2019eav, Chia:2022udn}, as it could spontaneously form bosonic condensates around the black holes and produce a wealth of interesting astrophysical signatures~\cite{Yoshino:2012kn, Yoshino:2014, Baryakhtar:2017ngi, Baumann:2018vus, Baumann:2019ztm, Chia:2020dye, Baryakhtar:2020gao, Siemonsen:2022ivj, Baumann:2021fkf, Tomaselli:2024bdd}. Owing to the unique nature of the event horizon, any potential departures of the black hole tidal heating effects in extreme-mass-ratio systems have also been proposed as signs of modifications of General Relativity or the existence of heavy exotic compact objects in the Universe~\cite{Ryan:1995wh, Datta:2019euh, Datta:2019epe, Cardoso:2019rvt}.

\vskip 4pt

While the physical imprints of other finite-size effects, such as the spin-induced moments~\cite{Poisson:1997ha,  Marsat:2014xea, Levi:2014gsa, Krishnendu:2017shb, Krishnendu:2018nqa,Chia:2020psj, Chia:2022rwc, Lyu:2023zxv} and the tidal deformability~\cite{love, Goldberger:2004jt, Flanagan:2007ix, Li:2007qu, Damour:2009vw, Binnington:2009bb, Vines:2011ud, Cardoso:2017cfl}, of a general body  on GWs have been studied extensively in the literature, the same cannot be said for tidal heating. This relative lack of progress stems primarily from the absence of a rigorous, first-principle derivation of this effect for a general astrophysical body on GW observables. To date, studies on tidal heating for a general body are either restricted to the Newtonian limit~\cite{Ripley:2023qxo} or adopt a phenomenological approach~\cite{Maselli:2017cmm, Datta:2019euh, Datta:2019epe, Datta:2020gem, Datta:2024vll}, whereby the GW phases and amplitudes are artificially deformed at the PN orders at which tidal dissipation are known to appear. In the latter case, ad-hoc parameters have been introduced to model tidal dissipation at 2.5PN and 3.5PN orders for rotating bodies and at 4PN for non-rotating bodies. Such parameterizations, while enabling some examination of this effect on GW observables, cannot be directly mapped to physical quantities such as the energy flux absorbed by the body. Moreover, to the best of our knowledge, most studies involving tidal heating in the context of data analysis have only been performed at the level of Fisher analyses of waveform mismatch~\cite{Maselli:2017cmm, Datta:2019euh, Datta:2019epe, Datta:2020gem, Datta:2024vll} but have not been applied to real data for parameter estimations studies (though see~\cite{Ripley:2023lsq} for recent result on tidal heating constraints the binary neutron star GW170817).

\vskip 4pt

In this paper, we expand upon previous investigations on tidal heating on both the theoretical frontier and from a data analytical perspective. On the theoretical front, we present a rigorous modeling of tidal dissipation of a single body by adopting the worldline effective field theory (EFT) formalism~\cite{Goldberger:2004jt, Goldberger:2005cd, Porto:2005ac,Porto:2007qi, Goldberger:2020fot, Charalambous:2021mea, Ivanov:2022hlo, Ivanov:2022qqt,Saketh:2022xjb} (see~\cite{Porto:2016pyg,Levi:2018nxp,Goldberger:2022ebt,Goldberger:2022rqf} for recent reviews). In the EFT framework, 
tidal dissipation 
can be described by a set of free 
coefficients in the non-local 
part of the retarded tidal response function
of a compact body. 
These coefficients appear 
in front of operators whose
structure is fully dictated by 
symmetries. 
EFT offers a systematic approach towards modeling both the spin-dependent and spin-independent tidal dissipation effects. Furthermore, this formalism \textit{a priori} makes no assumption about the nature of the binary components and is therefore applicable to all types of astrophysical bodies. 

\vskip 4pt

With the EFT, physically-motivated parameterizations of tidal heating for a general binary system can be derived in a straight forward manner. For binary systems with non-spinning components, we find that the leading-order tidal dissipation coefficient is described by the mass-weighted parameter $ \mathcal{H}_0$ -- see (\ref{eqn:H0}) for its definition -- and first appears at 4PN order in waveform observables.
$ \mathcal{H}_0$ can be thought of as a 
generalization of the Newtonian 
viscous lag time scale.
It is worth emphasizing that $ \mathcal{H}_0$ is the dissipative counterpart of the more well-known effective tidal Love parameter describing the conservative tidal deformation, which is often denoted by $\tilde{\Lambda}$ in the literature~\cite{Flanagan:2007ix, Favata:2013rwa}.\footnote{The tidal Love numbers have recently been a subject of intense theoretical research, 
especially in the context of black holes,
see e.g.~\cite{Damour:2009vw,Binnington:2009bb,Kol:2011vg,Hui:2020xxx,LeTiec:2020spy,Chia:2020yla,Charalambous:2021mea,Charalambous:2021kcz,Hui:2021vcv,Charalambous:2022rre,Ivanov:2022qqt,Hui:2022vbh,DeLuca:2022tkm,Charalambous:2023jgq,Rodriguez:2023xjd,DeLuca:2023mio,Charalambous:2024tdj,Berti:2024moe}.} 
For binary systems with spinning components, the leading-order tidal heating parameters first appear at 2.5PN order, and they are defined in (\ref{eqn:HsHa_linear}). Physically, 
they measure the body's angular
velocity. 
In the EFT, their origin can be traced back to the 
transformation from the co-rotating frame
to the local inertial frame. Thus, EFT automatically
incorporates possible 
radiation enhancement effects such as
superradiance. 
While it is well known that the spin-independent and spin-dependent tidal heating terms first appear at 4PN and 2.5PN orders respectively, our derivation provides a first-principles derivation of the functional form of the physically-motivated effective dissipation numbers in binary systems. 
Since the effective dissipation numbers described in this paper can be directly mapped to the microscopic properties of orbiting bodies, their measurements or constraints thereof provide meaningful physical interpretations on the nature of the binary sources. 

\vskip 4pt

In addition to the theoretical investigation of tidal heating, we also present the parameter estimation (PE) constraints on the dissipation numbers using the public LVK data. Crucially, we present our constrains for both binary black holes and for exotic compact objects --- in the former case, we exploit various properties which are unique to black holes, such as the black hole electric-magnetic duality~\cite{teukolsky1973perturbations,teukolsky1974perturbations,Chandrasekhar:1985kt,Goldberger:2005cd,Porto:2007qi, Chia:2020yla, Hui:2020xxx}, to obtain stronger PE constraints; while in the latter we make no such assumptions and therefore constrain all of the different dissipation numbers simultaneously in the PE process. 
Interestingly, although the spin-dependent effective numbers (\ref{eqn:HsHa_linear}), and the higher order terms thereof, formally appear at lower PN orders compared to the 4PN spin-independent coefficient (\ref{eqn:H0}), we found that the spin-independent coefficient $\mathcal{H}_0$ is best constrained as the spins of most detected BBHs are not very precisely measured and are largely consistent with zero~\cite{LIGOScientific:2020kqk, Roulet:2021hcu, KAGRA:2021duu}. While the current constraints on $\mathcal{H}_0$ for black holes is approximately two orders of magnitude larger than the theoretical prediction from General Relativity, future GW observations will rapidly improve the constraints as the detectors' sensitivity improves and the number of binary detection increases. All in all, this work presents the first rigorous derivation of tidal heating effects on GW observables using the EFT approach and provides the first physically-motivated constraints of tidal heating of GW sources.

\vskip 20pt

\noindent \textbf{Outline}: In Section~\ref{sec:summary} we present a summary and the main results of this work. In Section~\ref{sec:theory} we study the theoretical aspects of tidal dissipation for a general rotating body and its impact on GW observables of binary systems. In Section~\ref{sec:observations} we apply our theoretical results to the public LIGO-Virgo-KAGRA data and constrain the various dissipation numbers of the detected binary systems. Here we separate our analyses into two parts: the first assumes that the binaries are binary black holes, while the second relaxes this assumption and therefore applies to all kinds of exotic compact objects.  We present our conclusions and outlook in Section~\ref{sec:conclusions}. Details of the derivations for the waveform observables are shown in Appendix~\ref{appendix:derivation}.

\vskip 8pt

\noindent \textbf{Notations and Conventions}:  We use the superscript ${}^{\rm src}$ to distinguish source-frame and detector-frame quantities. For instance, the source-frame chirp mass is denoted by $\mathcal{M}^{\rm src}$ whereas the detector-frame chirp mass is $\mathcal{M}$. 
Latin letters $\{i, j, k\}$ denote spatial indices. We use $m_\ell$ to denote the azimuthal angular momentum number in order to avoid confusion with the component mass $m$. We adopt the following conventions for several convenient mass and spin quantities:
\begin{equation}
    \begin{aligned}
M & :=m_1+m_2 \, \qquad 
\eta  :=m_1 m_2 / M^2 \, \qquad
\delta  :=\left(m_1-m_2\right) / M \\
\chi_i & :=\mathbf{S}_i / m_i^2 \, \qquad
\boldsymbol{\chi}_s  :=\left(\boldsymbol{\chi}_1+\boldsymbol{\chi}_2\right) / 2 \, \qquad
\boldsymbol{\chi}_a  :=\left(\boldsymbol{\chi}_1-\boldsymbol{\chi}_2\right) / 2 
\end{aligned}
\end{equation}
where $\mathbf{S}_i$ is the component spin angular momentum and $\chi_i$ is the dimensionless spin.

\section{Summary of Main Results} \label{sec:summary}

Tidal effects probe the 
internal structure of a compact 
astrophysical body 
beyond the point particle 
approximation. 
In the post-Newtonian (PN) regime,
the leading order tidal effect 
on the gravitational waveforms 
stems from tidal heating (4PN for non-rotating bodies
and 2.5PN for spinning ones). 
The LVK GW data is usually analyzed under the ``standard physics'' 
assumption that 
detection triggers whose binary component masses are above $\approx 3M_\odot$
correspond to black hole-black hole mergers 
in general relativity.
Based on the PN counting, deviations from this scenario, 
if any, are expected 
to first show up at the level of 
tidal heating parameters. 
In this work, we demonstrate how 
to use these leading order effects
to test the nature of compact 
relativistic objects with GW data. 
Our work is a natural extension 
of the efforts to probe the nature of 
astrophysical 
compact objects with conservative 
tidal deformations, parameterized 
by static Love numbers. 
The effects of Love numbers,
however, 
first appear at 5PN, and hence we expect the 
tidal heating effects to provide
a stronger test of new physics from a PN counting perspective. 
Note that the Love numbers 
of exotic compact objects have been constrained with LVK data earlier in~\cite{Chia:2023tle}.

We derive the constraints on tidal heating of black holes and black hole-like exotic compact objects 
using the LIGO-Virgo O1-O3 
gravitational wave merger catalogs~\cite{LIGOScientific:2018mvr, LIGOScientific:2020ibl, LIGOScientific:2021usb, LIGOScientific:2021djp, Venumadhav:2019tad, Venumadhav:2019lyq, Olsen:2022pin, Nitz:2018imz, Nitz:2019hdf, Nitz:2021uxj, Nitz:2021zwj, Chia:2023tle, Mehta:2023zlk, Wadekar:2023gea}. 
To this end, we carry out two types of analyses. 
In our baseline analysis 
we place bounds on tidal 
heating of black holes, assuming 
the electric-magnetic duality that is
satisfied by black holes in four 
dimensional general relativity~\cite{teukolsky1973perturbations,teukolsky1974perturbations,Chandrasekhar:1985kt,Goldberger:2005cd,Porto:2007qi, Chia:2020yla, Hui:2020xxx}. 
This 
analysis effectively tests
whether dissipation
properties of black holes 
in the LVK data 
is consistent 
with general relativity. 
In the second analysis, 
for each LVK event normally
interpreted as a black hole - black hole merger, 
we derive bounds on a generic set of 
tidal heating parameters of the binary components.
This 
effectively 
constrains exotic compact objects
whose tidal properties are different from those of black holes.

We use the worldline effective field theory
(EFT)~\cite{Goldberger:2004jt, Goldberger:2005cd, Porto:2005ac,Porto:2007qi, Goldberger:2020fot, Charalambous:2021mea, Ivanov:2022hlo, Ivanov:2022qqt,Saketh:2022xjb} to describe effects of tidal heating
and dissipation on the gravitational waveforms. 
EFT allows one to parametrize these effects 
in a model-independent fashion, using symmetry
arguments only. We show that at the first non-trivial 
order in spin and frequency, these effects are 
captured 
by two free parameters, which can be thought of as generalizations 
of the tidal Love numbers. These parameters are defined as follows.

\begin{figure}[h!]
 \begin{adjustbox}{valign=c}
    \includegraphics[scale=0.4]{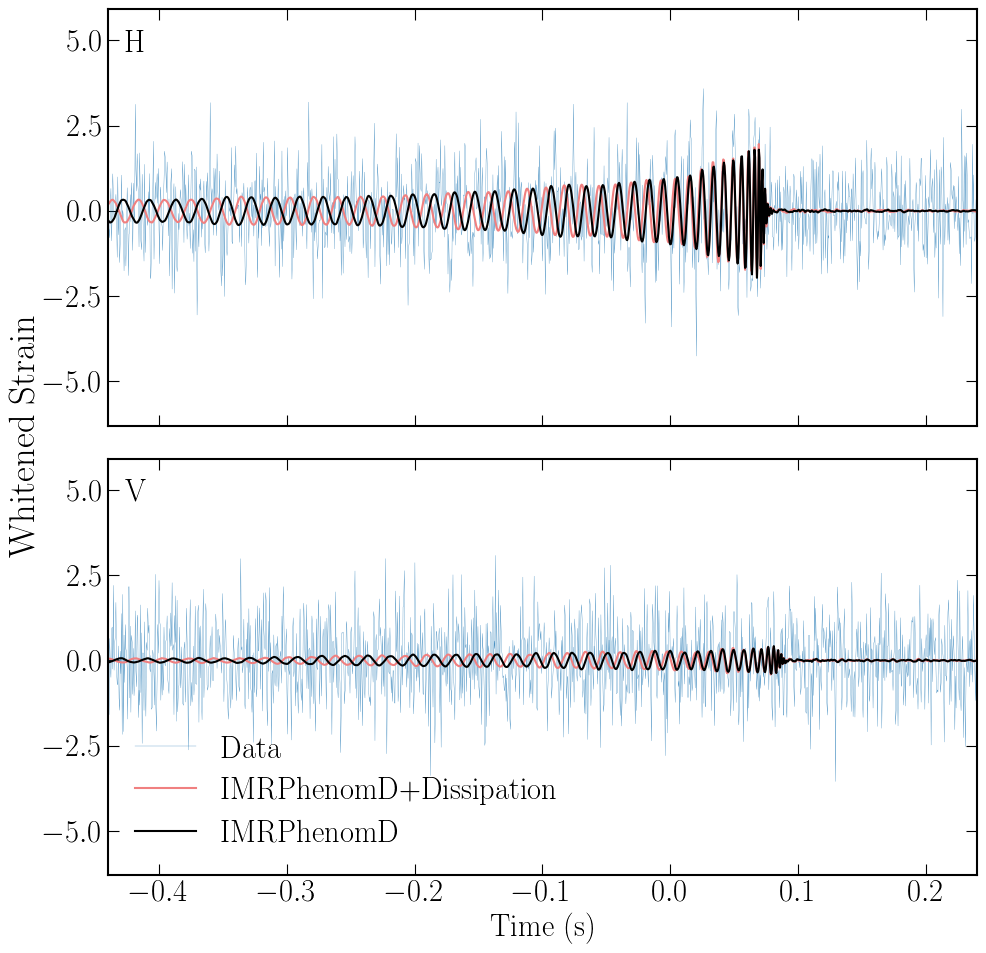}
     \end{adjustbox} 
\begin{minipage}{0.8\textwidth}
  \begin{subfigure}{\textwidth}
\includegraphics[width=0.49\textwidth]{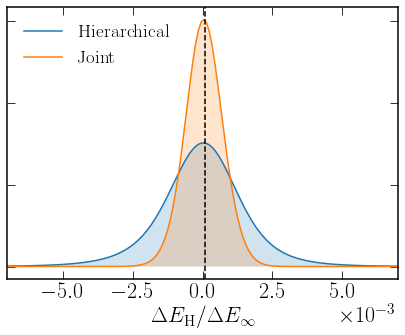}\vfill
\includegraphics[width=0.49\textwidth]{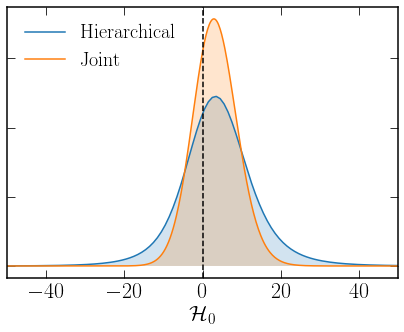}
 \end{subfigure}
 \end{minipage}
  \hspace{-6.9cm}
     \caption{
     \label{fig:summary_BBH}
    \textit{Left panel:}  GW strains of the IMRPhenomD and IMRPhenomD+Dissipation waveforms for a GW191216\_213338 event in the Hanford and Virgo data. We choose the individual 
    dissipation
    parameters $H_{1\omega} = H_{2\omega}=10$ to clearly illustrate dephasing of the waveform due to tidal heating.
  \textit{Upper right panel:} 
    Constraints on the ratio of the energy loss $\Delta E_{\rm H}/ \Delta E_{\infty}$ due to tidal dissipation relative to radiative loss at infinity. The black dashed line represents the GR prediction $\Delta E_{\rm H} / \Delta E_{\infty} \simeq ( 10^{-5},10^{-4})$. 
      \textit{Lower right panel:}  Constraints on 
      the spin-independent dissipation coefficient $\mathcal{H}_0$. 
      The black dashed line shows the GR prediction for equal mass binary black holes with $\mathcal{H}_0=2/45$.
    } 
\end{figure}

In Newtonian physics, a spherically symmetric isolated compact body subjected to 
a weak external qudrupolar tidal moment $E^{ab}$ 
generates the induced quadrupole 
moment $ Q_{ab}$, whose expansion in time derivatives takes the form~\cite{Poisson:2009di,poisson_will_2014}:
\be 
 Q_{ab} =
-\frac{2}{3G} k_2R^5 \[  E_{ab} - \mathcal{T}  \frac{d}{dt}E_{ab} + \cdots \] \,,
\ee 
where $R$ is the body's radius,
and $G$ is Newton's gravitational constant. 
The coefficient $k_2$ above 
is a Newtonian Love number, while $\mathcal{T} $ 
is the viscous time 
lag between the tidal field
and the body's response. 
In the presence of 
viscosity, 
the mass transfer rate due to the
absorption of gravitational radiation is given by~\cite{DEath:1975jps,Poisson:2004cw,Goldberger:2004jt,Goldberger:2005cd,Poisson:2009di,Goldberger:2020fot}, \be \label{eq:massdis0}
     \frac{d m}{dt} = \frac{2}{3G}R^5k_2\mathcal{T} 
    \dot{E}^{ab} \dot{E}_{ab}\,.
\ee 
Assuming that the body is slowly 
spinning, one can see that
viscosity 
also leads to the loss of the angular momentum $J$, i.e. the 
tidal field induces a torque~\cite{DEath:1975jps,poisson_will_2014}
\be 
   \frac{dJ}{dt}=-\frac{4}{3G}R^5k_2\mathcal{T}   \epsilon_{abc} \dot{E}^{a}_d{E}^{db}\hat s^c\,,
\ee
where $\hat s^c=(0,0,1)$ is a unit vector in the spin direction. 
EFT allows us to generalize these concepts beyond
the Newtonian limit, for a generic 
compact body, arbitrary spin,
and in the presence of relativistic 
non-linearity. In the observationally relevant case of 
small spin, the general expression 
for the dissipative (i.e. odd under time reversal) part of the electric-type induced 
quadrupole moment in the 
local asymptotic rest frame is given by 
\be 
\label{eq:indQ}
  Q^{ij} \Big|_{\rm dis.}
=   m (G m)^4 \[ (G m) H_{\omega} \frac{D}{D\tau} E^{ij} -H_{S} \chi \hat{S}^{\langle i}{ }_{ k} E^{k| j\rangle }  \]\,,
\ee 
where $\tau$ is proper time of the black hole's worldline, $\chi=J/(Gm^2)$
is the dimensionless spin, $\hat S^{ij}$ is the unit spin tensor. 
The free coefficient 
$H_{\omega}$ generalizes the viscous lag parameter,
while $H_{S}$ is a new coefficient that measures 
the absorption of mass and angular momentum due to
spin. Physically, it originates 
from a coordinate transformation
from the local rotating frame to the
local asymptotic inertial rest frame.
Essentially, this terms captures
superradiance and tidal locking.
The amplitude of $H_{S}$
is set by body's angular 
velocity measured at infinity.

\bgroup
\begin{table}[t!]
    \centering
    \setlength\tabcolsep{15pt}
    {\tabulinesep=1.0mm
    \begin{tabular}{|c|c|c|}
        \hline 
         \multirow{2}[0]{*}{} & \multirow{2}[0]{*}{Physical Quantity} & \multirow{2}[0]{*}{$90\%$ credible interval} \\
          & & \\
        \hline
\multirow{3}[0]{*}{Dissipation coefficients} & $\mathcal{H}_0$ & $-13 < \mathcal{H}_0 < 20$ \\
& $\mathcal{H}_1$  & $-18 < \mathcal{H}_1 < 11$  \\
& $\overline{\mathcal{H}}_1$ & $-18 < \overline{\mathcal{H}}_1 < 16$ \\
\hline
Relative energy loss & $\Delta E_{\rm H}/ \Delta E_{\infty}$ & $-0.0026 < \Delta E_{\rm H} / \Delta E_{\infty} < 0.0025$ \\
\hline
     \end{tabular}}
    \caption{Summary of the constraints on the tidal heating coefficients for the black hole population in the LIGO-Virgo O1-O3 catalog. We also provide constraints on the ratio of the energy loss to the black hole horizon due to tidal heating to the gravitational wave energy propagated towards infinity. The GR prediction is $\mathcal{H}_0=2/45$ for equal mass binary black holes and $\Delta E_{\rm H} / \Delta E_{\infty} \simeq (10^{-5}, 10^{-4})$. }
    \label{tab:pn_counting}
\end{table}
\egroup

$H_{S}$ and $H_{\omega}$ are the 
main two parameters that we constrain 
with the LIGO-Virgo-KAGRA data. 
To do so, we consistently derive the 
waveform modifications 
following from the induced 
quadrupole moment~\eqref{eq:indQ}.
$H_{S}$ and $H_{\omega}$ affect the waveforms at 2.5PN and 4PN,
respectively. 
We also study ``magnetic'' (parity-odd) tidal effects, 
which appear at a higher post-Newtonian (PN) order. 
In addition, we constrain a more 
intuitive physical quantity, the 
absorption of mass following from 
the generalized Eq.~\eqref{eq:massdis0},
\be 
 \label{eqn: mdot_single}
    \dot{m} =  m(Gm)^5 H_{\omega} \dot{E}^{ij} \dot{E}_{ij}- m(G m)^4 H_{S} \chi \( \dot E^{ij} E_{i}{}^k \hat{S}_{jk}\)  +\text{magnetic}  \,.
\ee 
In the upper right panel of Fig.~\ref{fig:summary_BBH}, we present the population level constraint on the ratio between the energy lost due to tidal dissipation, $\Delta E_{\rm H}$, and the radiative energy at infinity, $\Delta E_{\infty}$.  We establish that, at the population level, the constraint falls within the range of $-0.0026 < \Delta E_{\rm H} / \Delta E_{\infty} <0.0025$, with a $90\%$ credible level, as determined through a hierarchical Bayesian approach. 
Note that this constraint includes the 
possibility of the radiation enhancement,
similar to superradiance, which corresponds to 
negative values of $H_\omega$.
Our result is 
consistent with 
the general relativity 
prediction that BBHs should lose about a $10^{-5}-10^{-4}$ fraction of 
the total 
radiated energy to tidal dissipation.
 Our main constraints are summarized in Table~\ref{tab:pn_counting}.

The primary factor contributing to tidal dissipation is represented by the 4PN spin-independent
dissipation parameters $H_{\omega}$. 
At leading order, it affects the waveforms 
through the mass weighted combination $\mathcal{H}_0$,
\be
    \mathcal{H}_0 \equiv \frac{1}{M^4}  \left( m_1^4 H_{1 \omega} + m_2^4 H_{2\omega} \right)  \,,\label{eqn:H0}
\ee
where $m_{1,2}$ are the component masses, $M=m_1+m_2$,
and $H_{1,2\omega}$ are individual dissipation numbers
of binary components. 
We show the typical 
waveform modifications due to this parameter
in the left panel of Fig.~\ref{fig:summary_BBH}.
In the lower right panel of Fig.~\ref{fig:summary_BBH}, we display the combined posterior, 
with the constraint $-13 < \mathcal{H}_0 < 20$ at $90\%$ CL.

For binaries with spinning components, the leading-order tidal heating effects first appear at 2.5PN and are quantified by the symmetric and anti-symmetric combinations
\be 
 \mathcal{H}_1 \equiv \frac{1}{M^3} \left( m_1^3 H_{1S} + m_2^3 H_{2S} \right) \,, \quad \overline{\mathcal{H}}_1 \equiv \frac{1}{M^3} \left( m_1^3 H_{1S} - m_2^3 H_{2S} \right) \,,  \label{eqn:HsHa_linear}
\ee 
where $H_{1,2S}$ are the linear-in-spin tidal dissipation numbers of the binary components.
The physical imprints associated 
with $\mathcal{H}_1$ and $\overline{\mathcal{H}}_1$ on GW observables are linearly proportional to the symmetric and anti-symmetric combinations of the component spins. 
Since most of the LVK
mergers have small spin,
$\mathcal{H}_1$ and $\overline{\mathcal{H}}_1$
are somewhat less constrained
than $\mathcal{H}_0$,
even though the effects of these 
parameters is nominally
stronger in the PN counting. 
Our constraints on these parameters
from the LVK data are displayed in Table~\ref{tab:pn_counting}.

We discuss other subdominant effective dissipation numbers, such as the cubic-in-spin dissipation numbers and the magnetic counterparts of (\ref{eqn:H0}) and (\ref{eqn:HsHa_linear}), in Section~\ref{sec:theory}. 
Constraints on the magnetic-type 
dissipation coefficients are
presented in Section~\ref{sec:observations}.

\begin{figure*}[t!]
    \centering
    \hspace{100pt}
    \begin{adjustbox}{valign=t}
    \includegraphics[width=0.70\textwidth, trim=0 10 0 0]{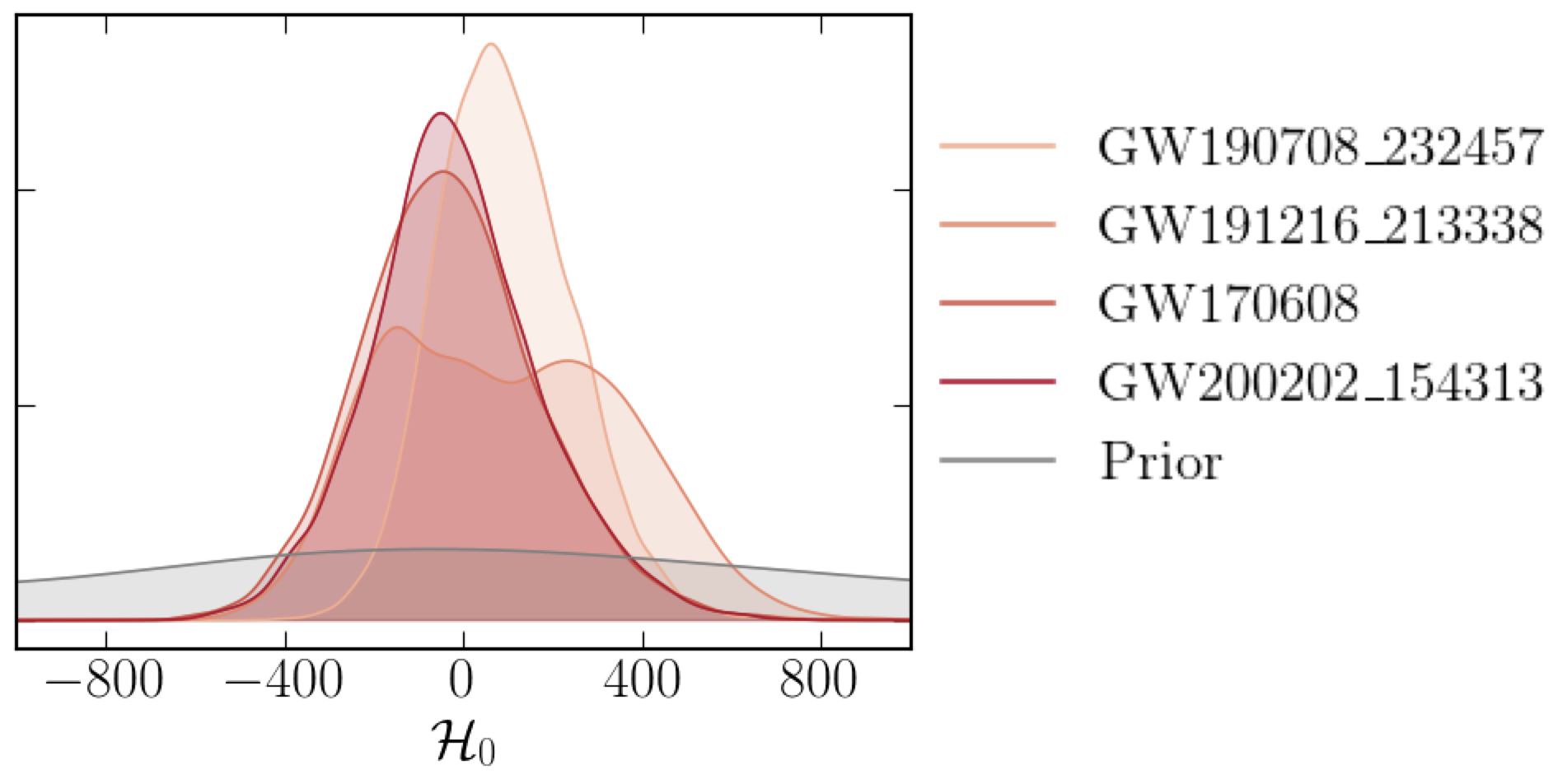}
    \end{adjustbox}
    \\
    \hspace{-30pt}
    \begin{adjustbox}{valign=t}
    \includegraphics[width=0.42\textwidth, trim=0 10 0 0]{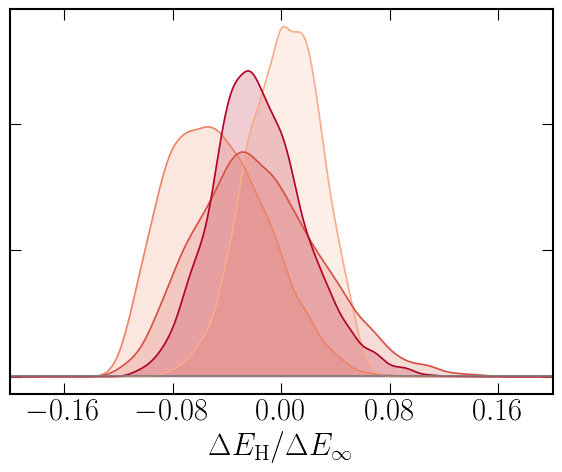}
    \end{adjustbox}
    \caption{ ({\it top}) Inspiral-only constraints on the dissipation number $\mathcal{H}_0$ for selected events. ({\it bottom}) Inspiral-only constraints on the ratio of the energy loss $\Delta E_{\rm H}/ \Delta E_{\infty}$.}
    \label{fig:summary_ins_only}
\end{figure*}

From the technical side, 
it is worth stressing 
that we use exclusively the inspiral phase in our analysis
in order to 
be conservative and agnostic
about the nature of black-hole like compact objects. 
To achieve this, we truncate both the data, denoted as $d$, and the waveform, $h$, at the so-called tapering frequency, defined as $f_{22}^{\rm tape} = 0.35 f_{22}^{\rm peak}$, with $f_{22}^{\rm peak}$ being the peak frequency. This methodology allows us not to presuppose that the signals originate from black holes, thereby making our findings potentially applicable to a broader range of exotic binary systems. In Fig.~\ref{fig:summary_ins_only}, we present our constraints on the dominant dissipation number, $\mathcal{H}_0$, and the ratio of energy loss, $\Delta E_{\rm H} / \Delta E_{\infty}$, for a few selected events. Our results indicate that for individual events $|\mathcal{H}_0| \lesssim 400$ and $|\Delta E_{\rm H} / \Delta E_{\infty}| \lesssim 0.08$ within a $90\%$ credible interval.

\section{Tidal Heating Effects on Gravitational Waves: Theory}
\label{sec:theory}

In this section, 
we review the description of 
tidal dissipateve dynamics of a single rotating body from the perspective of the worldline EFT formalism~\cite{Goldberger:2004jt, Goldberger:2005cd, Porto:2007qi, Porto:2016pyg, Goldberger:2020fot,Charalambous:2021mea,Ivanov:2022hlo,Ivanov:2022qqt,Saketh:2022xjb}. Then in \S\ref{sec:gw_observables}, we discuss tidal dissipation in the context of binary systems and present results for the imprints of this effect on the GW phase and amplitude.

Most of the material presented in 
\S\ref{sec:single_body} and \S\ref{sec:heatrot}  
is well known in the literature; we refer the reader 
to the original work and excellent reviews~\cite{Goldberger:2005cd,Porto:2005ac,Porto:2007qi,Goldberger:2020fot,Goldberger:2022ebt,Goldberger:2022rqf}
for further details. Our goal here is to 
highlight the main conceptual 
features of EFT that are relevant
in the context of LVK data analysis. 
In particular, 
we would like to point out 
that EFT makes manifest the 
connection between 
terms that appear in
the LVK waveforms and the on-shell
amplitudes for gravitational
scattering computed in~\cite{Goldberger:2005cd,Porto:2005ac,Porto:2007qi,Goldberger:2020fot,Saketh:2023bul,Ivanov:2024sds}.

\subsection{Worldline EFT for Tidal Heating of a Non-Rotating Body} \label{sec:single_body}

In worldline EFT, compact bodies are described, at leading 
order, as point particles. In the simplest example
of a body without spin, the leading order EFT action 
for a single body 
is given by 
\be 
\label{eq:pp0}
S^{(0)} = -m\int ds \,,
\ee 
where m is body's mass and $s$ is the worldline 
parameter which we will often choose to 
be equal to body's proper time $\tau$.
In addition to action~\eqref{eq:pp0}
one should consider the 
perturbative general relativity action
for ``bulk gravity,''
\be 
S_{\rm GR} = -\int d^4x~\sqrt{g}\frac{R}{16\pi G}\,.
\ee 
Conservative dynamical effects beyond the point particle approximation,
such as tidal deformations, are described 
by higher order operators coupled to the worldline, 
\be 
S^{(2)}= mR^4 \int ds (c_E E^{\mu\nu}E_{\mu\nu} +
c_B B^{\mu\nu}B_{\mu\nu}) + ...\,,
\ee 
where we used the electric 
and magnetic parts of the Weyl tensor $C_{\mu\nu\alpha \beta}$,
\be 
E_{\mu\nu} = C_{\mu\rho\nu\sigma}u^\rho u^\sigma\,,\quad 
B_{\mu\nu} = \frac{1}{2}
\varepsilon_{\mu\rho\alpha \beta}C^{\alpha\beta}{}_{\nu\sigma}u^\rho u^\sigma\,,
\ee 
$u^\mu$ is
the four-velocity of the point particle.
Parameters $c_{E,B}$ are dimensionless EFT Wilson coefficients (counterterms),
which provide a gauge invariant definition  
for the static Love numbers.
Note that the formal expansion parameter
in worldline EFT for compact bodies
is $\omega R\ll 1$, i.e. it is applicable 
only to describe dynamics on scales much larger 
than the size of the compact body,
which is appropriate for the 
binary dynamics where $\omega$
can be assosiated with 
the orbital frequency. In what follows, 
in the context of 
non-rotating bodies, 
it will be convenient to work 
with tetrads $e^a_\mu$ 
defining an orthonormal frame,
which satisfy 
\be 
    \eta_{a b} e_\mu^a e_\nu^b=g_{\mu \nu}(x), \quad g_{\mu \nu}(x) e_a^\mu e_b^\nu=\eta_{a b}\,.
\ee
In the body's rest frame, these tetrads reduce to 
\be
e^\mu_a = \delta^\mu_a \,,
\ee
where 
$a = 0, 1,2,3$ denote 
local 
Lorentz indices. 
In order to describe dissipative effects, 
one has to introduce 
composite multipole moments $Q_{ab}$
that are coupled to 
the external gravitational field~\cite{Goldberger:2005cd,Goldberger:2020fot}, 
\be 
\label{eq:multi_tides}
S = -\int ds \, Q_{ab}^E E^{ab} - \int ds \, Q_{ab}^B B^{ab}~\,.
\ee 
The explicit form of the multipole 
moments $Q_{ab}^{E/B}$
is unknown. Classical physical 
observables depend only on their 
retarded
Green's functions. (See Refs.~\cite{Goldberger:2019sya,Goldberger:2020wbx,Goldberger:2020geb} for the treatment of quantum effects.) These can be matched 
from gauge-invariant observables 
such as the cross-section
for the compact body to absorb
gravitons. For instance,
a leading order EFT calculation gives~\cite{Goldberger:2005cd}
\be 
\sigma_{\rm abs}(\omega)=G\pi \omega^3 
\varepsilon^*_{ab}\varepsilon_{cd}\langle Q^{ab}_E Q^{cd}_E\rangle(\omega)+\text{magnetic}\,,
\ee 
where $\langle Q^{ab}_E Q^{cd}_E\rangle$
is the Wightman correlator.
The underlying spherical symmetry of the 
unperturbed compact body dictates that
\be 
\langle Q^E_{ab} Q^E_{cd}\rangle(\omega) \equiv \int ds e^{i\omega s}\langle Q^E_{ab}(s)Q^E_{cd}(0)\rangle = \frac{1}{2}\left(
\eta^\perp_{ac}\eta^\perp_{bd}
+\eta^\perp_{ab}\eta^\perp_{bc}
-\frac{2}{3}\eta^\perp_{ab}\eta^\perp_{cd}
\right)F_E(\omega)\equiv \mathcal{P}^{(0)}_{ab,cd} F_E(\omega)\,,
\ee 
for some scalar function $F_E(\omega)$
(and the same for $B$),
where $\eta^\perp_{ab} = \eta_{ab}-u_a u_b$.
is the metric in the compact body's frame. 
Comparing above with the cross-section
calculated in GR allows one to 
determine the functional form 
of $F_{E/B}(\omega)$. For instance, matching 
to the graviton absorption cross-section
in black hole perturbation theory one obtains~\cite{Goldberger:2005cd} 
\be 
F_E(\omega) = F_B(\omega) =\Theta_H(\omega)\frac{2 r_s^6\omega}{45G}\,,
\ee 
where $\Theta_H(\omega)$ denotes the Heaviside 
step function.
After matching, this expression can be used to make further
prediction for other observables.
In the context of gravitational wave astronomy, 
one is interested in the expectation
values $\langle Q_{ab}^{E/B}\rangle$
over the ensemble of small-scale fluctuations
produced by the internal degrees of freedom
of a compact body. In this case, 
for the adiabatic phase of the inspiral, 
one can assume that the external gravitational
fields are weak, and apply linear 
response theory, yielding  
\be 
\langle Q_{ab}^{E}\rangle = \int ds' G^{\rm ret}_{ab,cd}(s -s') E^{cd}(x(s'))\,, 
\ee 
and the same for $Q_{ab}^B$.
Note that $E^{cd}(x(s'))$ above 
is the external tidal 
field
evaluated on the compact body's worldline. The 
Green's function above cannot be calcualted in
general. Its local part is indistingushable 
from the contributions from the local 
counterterms in $S^{(2)}$. The non-local 
contribution, however,
is calculable, and 
can be
expressed through the correlators
that we have matched before
from the absorption cross-section, yielding e.g. for the electric part~\cite{Goldberger:2020fot}
\be 
\langle Q_{ab}^{E}\rangle = \frac{4r_s^2(Gm)^4}{45G}
\mathcal{P}^{(0)}_{ab,cd}\frac{d}{ds}E^{cd}(x(\tau))=
\frac{4r_s^2(Gm)^4}{45G}
\frac{d}{ds}E^{ab}(x(\tau))\,.
\ee 
In the first 
approximation, i.e.
neglecting  
interactions of the 
internal multipoles with
``bulk'' gravitons, 
EFT 
gives the same result as the 
classical linear response theory post-Newtonian adiabatic 
expansion over time derivatives, 
\be 
\label{eq:response0}
\langle Q_{ab}^{E}\rangle = 
 m (Gm)^4 \[ c_0  E_{ab} + (Gm) h_0 \frac{d}{d\tau}E^{ab} + \cdots \] \,,
\ee 
where we inserted $m$ and $G$ in order to make the 
coefficients $c_0$ and $h_0$
dimensionless. In the 
traditional post-Newtonian context, 
$c_0$ is 
interpreted as a static Love number. In EFT, 
at first order, 
physical observables 
depend on the combination of $c_0$ and the local counterterms in $S^{(2)}$.

We stress that strictly speaking, 
Eq.~\eqref{eq:response0}
is true only
at the zeroth order
in the interaction 
with bulk gravitons. In general, 
tail effects make 
the relationship
between $Q_{ab}$
and $E_{ab}$ non-local. 
In the context of 
linear response theory,
this can be heuristically interpreted  
as a running of 
dissipation coefficients. In
the strict mathematical 
sense, within EFT, 
this corresponds to the 
renormalization
of the two-point function 
of $Q_{ab}$
due to its dressing with 
graviton loops~\cite{Ross:2012fc,Saketh:2023bul}. 

For a general body, the leading order tidal heating effect is captured by a single 
parameter $h_0$.
In the particular example of a black hole, matching of
the absorption cross-sections yield
\be 
h_0 = \frac{4r_s^2}{45G^2m^2}=\frac{16}{45}\,.
\ee 
To calculate non-conservative 
effects on the binary dynamics one has to introduce a 
composite momentum 
operator $p_a$
and conveniently 
rewrite the point particle 
action as 
\be 
S^{(0)} = -\int dx^\mu e_\mu^a p_a +\int ds~(p_a p^a-m^2)~\,.
\ee 
In the body's rest frame 
this action takes
to usual form 
of 
the point particle action.
Varying $S^{(0)}$
and $S^{(2)}$
w.r.t. $x^\mu$
one obtains,
in Fermi normal coordinates~\cite{Goldberger:2020fot}, 
\be 
\frac{Dp^\mu}{Ds}=
e^a_\rho e^b_\sigma
[\langle Q^E_{ab} \rangle \nabla^\mu E^{\rho\sigma}
+ 
\langle Q^B_{ab} \rangle \nabla^\mu B^{\rho\sigma}
]\,,
\ee 
where $p^\mu = p^a e^\mu_a$. Multiplying 
this by $p^\mu$ we arrive
at the final equation for the 
mass dynamics
\be 
\frac{dm^2}{ds} = 
2
e^a_\rho e^b_\sigma
[\langle Q^E_{ab} \rangle p_\mu\nabla^\mu E^{\rho\sigma}
+ 
\langle Q^B_{ab} \rangle p_\mu \nabla^\mu B^{\rho\sigma}
]\,.
\ee
If the response is purely 
conservative, $\langle Q_{ab}\rangle \propto E_{ab}$, the above equation
integrates to a trivial solution. 
The dissipative part 
of the response at the leading 
order gives 
\be 
\label{eq:mdot}
\frac{dm^2}{ds} = 
2(Gm)^5 m^2 h_0
[\dot E_{\rho\sigma}^2
+ 
\dot B_{\rho\sigma}^2
]\,. 
\ee
This equation can be used now to 
calculate fluxes relevant for 
gravitational waveforms. 
For black holes, the result is~\cite{Goldberger:2005cd}
\be 
\frac{dm}{ds} = 
\frac{16(mG)^5m}{45}
[\dot E_{\rho\sigma}^2
+ 
\dot B_{\rho\sigma}^2
]\,. 
\ee
which precisely matches the 
classical general relativity 
result for the absorption
of mass by a black hole 
in the adiabatic regime~\cite{Poisson:2004cw}.  
Eq.~\eqref{eq:mdot}
will be used as input
in gravitational waveform
calculations below. 

\subsection{Tidal Heating of a Rotating Compact Body}
\label{sec:heatrot}

The extension of the worldline EFT 
approach to spinning bodies 
is as follows. First, 
one writes down an 
effective action for a point particle with spin, 
\be 
\label{eqS0spin}
S^{(0)}=-\int dx^\mu e_\mu^a p_a +\int ds (p_ap^a-m^2)
+\frac{1}{2}\int d\lambda S^{ab}\Omega_{ab} + \int ds l_a S^{ab}p_b\,,
\ee 
with $ds=d\lambda e(\lambda)$,
and 
$\lambda$ denotes a
worldline's affine parameter. 
The action \eqref{eqS0spin} 
defines the spin tensor 
$S_{ab}$ as a canonical
conjugate of the angular
velocity. 
$l_a$ in the rightmost above term
is the Lagrange multiplier 
that eliminates 
spurious degrees of 
freedom in $S_{ab}$.
As before, $p_a$ and $S_{ab}$
are treated as composite operators 
that depend on 
unspecified 
internal dynamics of the 
compact body.
It will be convenient to define tetrades that describe 
a co-rotating frame, $e_a^0 = \delta^0_a$ and 
\be  
e^i_a = \begin{bmatrix}
    \cos(\Omega x^0) & -\sin(\Omega x^0) & 0 \\
    \sin(\Omega x^0) & \cos(\Omega x^0) & 0 \\
    0 & 0 & 1 \\
\end{bmatrix}\,,
\ee 
where $\Omega$ is 
body's angular velocity. The 
angular velocity tensor reads
\be 
    \Omega_{ab} \equiv - g_{\mu\nu} e_{a}^\mu \frac{D}{Ds} e_b^\nu = - \Omega_{ba} ~
\ee
Action \eqref{eqS0spin}
generates the usual Papapetrou-Mathison
equations on 
$p_\mu$ and $S_{\mu\nu}$
for a point particle. In order to
describe finite-size effects, 
we must supplement action \eqref{eqS0spin}
with the finite-size action \eqref{eq:multi_tides}. 
The generalized 
 Papapetrou-Mathison equations 
 for mass and spin dynamics then read~\cite{Goldberger:2020fot}
\be 
\begin{split}
& 
\frac{dm^2}{ds} = 
2
e^a_\rho e^b_\sigma
[\langle Q^E_{ab} \rangle p_\mu\nabla^\mu E^{\rho\sigma}
+ 
\langle Q^B_{ab} \rangle p_\mu \nabla^\mu B^{\rho\sigma}
]\,,\\
&
\frac{d S^2 }{ds} =
\langle Q^E_{ab} \rangle E^{bc}S^a_c
+ 
\langle Q^B_{ab} \rangle B^{bc}S^a_c
\,,
\end{split}
\ee 
where $S^2 = \frac{1}{2}S^{\mu\nu}S_{\mu\nu}$.
Note that the derivatives w.r.t. $s$
include both the intrinsic time-dependence 
of the body as well as the co-rotation, i.e.
\be \label{eq:co-rotation}
    \frac{D}{Ds} E^{ab} = e^{a}_\mu e^{b}_\nu \frac{D}{Ds} E^{\mu\nu} + \Omega^a{}_c E^{cb} + \Omega^{b}{}_c E^{ca} \,.
\ee
In order to extract the two-point functions of $Q^{E,B}_{ab}$ one can first use 
symmetry arguments to decompose 
$\langle Q^{E,B}_{ab} Q^{E,B}_{cd}\rangle$
over SO(3) STF tensors with different azimuthal numbers $m_\ell$, 
\be 
\langle Q^{E/B}_{ab} Q^{E/B}_{cd}\rangle
= \sum_{i=0}^4 A^{E/B}_i (s-s')
\mathcal{P}_{ab,cd}^{(i)}\,,
\ee 
where 
the projection tensors
$\mathcal{P}_{ab,cd}^{(i)}$
can be 
found in \cite{Goldberger:2020fot}. 
Then 
one can calculate the probability 
for a compact body to absorb a graviton
in the EFT, yielding
\be 
p(1\to 0) = \frac{4G\omega^5}{5}\sum_{i=0}^4 m_\ell^i \left(A^E_i(\omega-m\Omega)
+A^B_i(\omega-m\Omega)
\right)\,,
\ee 
and match it to general relativity results~\cite{Starobinsky:1973aij,Starobinskil:1974nkd}. 
For instance, for black holes 
one can obtain  
\be 
\begin{split}
& A_0^{E/B}(\omega) = \frac{16(r_+^2+a^2)}{45 G}(Gm)^4(1-\chi^2)^2 \Theta_H(\omega)\omega \equiv h_0 
\Theta_H(\omega)
\omega \,,\\
& A_2^{E/B}(\omega) = \frac{2(r_+^2+a^2)}{9 G}(Gm)^4\chi^2 (1-\chi^2) \Theta_H(\omega)\omega 
\equiv h_2 
\Theta_H(\omega)
\omega
\,,\\
& A_4^{E/B}(\omega) = \frac{8(r_+^2+a^2)}{45 G}(Gm)^4 \chi^4 \Theta_H(\omega)\omega \equiv h_4 
\Theta_H(\omega)
\omega\,,
\end{split}
\ee 
and $A_1^{E/B}=A_3^{E/B}=0$,
where we used $\chi = a/(Gm)$
and $r_+=M+\sqrt{M^2+a^2}$ is the outer horizon
of a rotating black hole.
We see that $A_i^{E/B}/\omega$
is constant for physical 
positive frequencies. 
Importantly, with our choice of tetrades, the supperradiance effect
is captured automatically.
The supperadiance 
appears naturally in the EFT 
approach as a result of a 
coordiante transformation
from a body's rest frame 
where the correlators of 
$Q_{ab}^{E/B}$ are defined, 
to local static frame of the
asymptotic observer 
where the measurements are 
done. 
Matching EFT and 
black hole perturbation
theory absorption cross-sections
we obtain 
\be 
\Omega  =\Omega_H = \frac{a}{r_+^2+a^2}\,.
\ee 
Using the Wightman two point functions of composite multipoles,
one can compute the 
non-local part of 
retarded Green's 
function and obtain 
\be 
\langle Q_{ab}^{E}(s)\rangle\Big|_{\rm non-local}
=\sum_{i=0}^4 h_i 
\mathcal{P}_{ab,cd}^{(i)}\frac{d}{ds}E^{cd}\,,
\ee 
with $h_1=h_3=0$,
and an analogous expression for 
the magnetic response.
We note, again, 
that at the leading order, 
the EFT expansion for the two-point
function
reproduces the classical expression for the linear
response 
of the quadrupole 
moments,

\be
\label{eq:response}
   \begin{split}
\langle Q_{ab}^E \rangle & = - m (Gm)^4 \[ (\lambda^E)_{abcd} E^{cd} - (Gm) (\lambda^E_{\omega})_{abcd} \frac{D}{Ds}E^{cd} + \cdots \] \, , \\ 
\langle Q_{ab}^B \rangle  & = - m (Gm)^4 \[ (\lambda^B)_{abcd} B^{cd} - (Gm) (\lambda_{\omega}^B)_{abcd} \frac{D}{Ds} B^{cd} + \cdots \] \, , 
\end{split}
\ee
where $(\lambda^{E/B})_{abcd}$ and $(\lambda^{E/B}_{\omega})_{abcd}$ 
are linear static and time-dependent response tensors 
in the body's rest frame, and dots stand for higher order time derivatives. 
The transformation from the local
co-rotating frame to the local inertial frame is given by Eq.~\eqref{eq:co-rotation}.

In what follows we will 
work within a linear response
approach, which is 
sufficient at the 
leading order in the PN scheme.  
We will use the symmetries of the problem, 
parity and axial symmetry, to expand the above
response tensors $(\lambda^{E/B})_{abcd}$ and $(\lambda^{E/B}_{\omega})_{abcd}$ over 
an appropriate basis of symmetric trace-free (STF)
tensors.
We will use spin to describe the breaking 
of rotational symmetry.
Specifically, we will use the unit spin tensor $\hat{S}_{ab}$ (with normalization $\hat{S}_{ab}\hat{S}^{ab} = 2$) and the Pauli-Lubanski unit spin vector, $\hat{s}^{a}\equiv (1/2)\epsilon^{abc}\hat{S}_{bc}$. 
The isomorphism between STF tensors and spherical 
harmonics can be used to connect 
tensors in our basis 
with the decomposition of the 
two-point functions of $Q$ in harmonic space that we used before.
Then, 
one can write the following 
expression for the frequency-independent part of the response:

\be 
\label{eqn:static_decomposition}
\begin{split}
            (\lambda^{E/B})^{ab}_{cd} = &\Lambda^{E/B} \hskip 1pt \delta_{\langle c}^{\langle a } \delta_{d \rangle}^{b \rangle}+ \Lambda_{S^2}^{E/B} \hskip 1pt  \chi^2 \hskip 1pt  \hat{s}^{\langle a} {s}_{\langle c} \delta_{d \rangle}^{b \rangle}  + \Lambda_{S^4}^{E/B} \hskip 1pt  \chi^4 \hskip 1pt \hat{s}^{\langle a} \hat{s}_{\langle c} \hat{s}^{b \rangle} \hat{s}_{d \rangle} \,,
            \end{split}
\ee
where angular brackets denote the operation of symmetrization and a subsequent subtraction of traces. 
The three response tensors above are degenerate with 
the contribution of local worldline counterterms.
This is because the response operators associated with tensors $\Lambda^{E/B}, \Lambda_{S^2}^{E/B}$, and $\Lambda_{S^4}^{E/B}$ are time-reversal even. 
Note that the
response tensor above could 
include extra terms,

\be 
\lambda^{E/B}{}^{ab}_{cd}\supset 
           \tilde H^{E/B}_{S}\chi \hat{S}^{\langle a}_{\langle c}\delta^{b\rangle }_{d\rangle } + \tilde H_{{S}^3}^{E/B} \hskip 1pt  \chi^3 \hskip 1pt \hat{s}^{\langle a} \hat{s}_{\langle c} \hat{S}^{b\rangle}{ }_{d \rangle}\,,
\ee 
which 
nominally describe static dissipation due to spin
in the co-rotating frame. At  the level of the Wightman function,
these correspond to 
$A^{E/B}_{1}$
and $A^{E/B}_{3}$.
They are not forbidden by symmetries, but they do not have a
classical interpretation in the 
Newtonian limit. 
In addition, these parameters 
vanish identically for black holes.  Therefore, 
in what follows we will 
not consider these
operators. 
 
A similar symmetry-based 
decomposition can be applied to the time-dependent response tensors in

\eqref{eq:response}:
\begin{equation}
\label{eqn:dynamic_decomposition}
        \begin{aligned}
            (\lambda_{ \omega}^{E/B})^{ab}_{cd} & = H_{ \omega}^{E/B} \hskip 1pt \delta_{\langle c}^{\langle a } \delta_{d \rangle}^{b \rangle}+ \Lambda_{{S}, \omega}^{E/B} \hskip 1pt  \chi \hskip 1pt \hat{S}^{\langle a}{ }_{\langle c} \delta_{d \rangle}^{b \rangle}+ H_{S^2,  \omega}^{E/B} \hskip 1pt  \chi^2 \hskip 1pt  \hat{s}^{\langle a} \hat{s}_{\langle c} \delta_{d \rangle}^{b \rangle} \\
            & \quad + \Lambda_{{S}^3,  \omega}^{E/B} \hskip 1pt  \chi^3 \hskip 1pt \hat{s}^{\langle a} \hat{s}_{\langle c} \hat{S}^{b\rangle}{ }_{d \rangle}+ H_{{S}^4,  \omega}^{E/B} \hskip 1pt \chi^4 \hskip 1pt  \hat{s}^{\langle a} \hat{s}_{\langle c} \hat{s}^{b \rangle} \hat{s}_{d \rangle}  \,.
        \end{aligned}
\end{equation}
The operators with even powers of spin, associated with $H_{\omega}^{E/B},H_{S^2, \omega}^{E/B}$, $H_{S^4,  \omega}^{E/B},$ have one time derivative 
and 
hence they are odd under time reversal, i.e. they capture dissipative effects. These operators
cannot be generated by a local worldline action. 
In contrast, the other terms above 
correspond to conservative time-reversal even operators, 
which be reproduced by  
local counterterms.

Let us consider now 
the term proportional 
to $H_{\omega}^{E/B}$
and focus, for concreteness on
the electric response,
\be 
\langle Q_{ab}^E\rangle\Big|_{\rm non-local}
=m(Gm)^5
H_{ \omega}^E
\frac{D}{Ds}E_{ab}\,.
\ee 
Using the analog of the Euler rotation equation~\eqref{eq:co-rotation},
and switching back asymptotically 
flat inertial frame 
we obtain, 
\be 
\label{eq:corot2}
\begin{split}
\langle 
Q^{ij}_{E}
\rangle\Big|_{\rm non-local} = m (G m)^5 
H_{S^0, \tilde \omega}^E\[ 2 \Omega^{\langle i}{ }_{ k} E^{k|j\rangle} + \frac{D}{D\tau} E^{ij} \] \,.
\end{split}
\ee 
In harmonic space, we obtain 
the following expression for the 
angular coefficients~\cite{Charalambous:2021mea}
\be 
\label{eq:harmonic_superrad}
    \langle Q_{\ell m_\ell}^E  \rangle\Big|_{\rm non-local} =-i m (G m )^5 H_{  \omega}^E (\omega - m_\ell \Omega) E_{\ell m_\ell} \,,
\ee
where $\Omega$ is the angular velocity 
of rotation.
We see that the square brackets in Eq.~\eqref{eq:corot2} describe the 
kinematic effect of superradiance
encoded
in the frame transformation.
The EFT expansion in the 
co-rotating frame incorporates this 
phenomenon automatically. 

Most of the O1-O3 LIGO-Virgo events 
have small spin~\cite{LIGOScientific:2021usb, LIGOScientific:2021djp, Roulet:2021hcu}, 
which suggests carrying out parameter
estimation only for spin-independent 
and linear in spin 
effects. If we consider 
only the time-derivative operators
in the co-rotating frame, we
end up with three free parameters,
$H_{\omega}^{E}, H_{ \omega}^{B}$
and $\Omega$
in the local asymptotic frame,
see Eq.~\eqref{eq:corot2},
\be 
\begin{split}
    \langle 
Q^{ij}_{E/B}
\rangle\Big|_{\rm non-local} =  m (G m)^4 \[
(G m) H_{\omega}^{E/B}\frac{D}{D\tau} E^{ij}
-H_{S}^{E/B} \chi \hskip 1pt \hat{S}^{\langle i}{ }_{ k}  E^{k|j\rangle }  \] \,,
\end{split}
\ee
with
$H^{E/B}_{S}=-{2Gm\Omega}{\chi}^{-1}H_{\omega}^{E/B}$. 
The generalized mass and spin dynamics
equations following from the symmetry-based ansatz~\eqref{eqn:dynamic_decomposition}
are given by 
\be 
    \begin{aligned}
    \dot{m} & =   m(G m)^4 \Bigg[ (Gm) H_{ \omega}^E \dot{E}^{ij} \dot{E}_{ij}
    +(Gm) H_{ \omega}^B \dot{B}^{ij} \dot{B}_{ij}
    -H_{S}^E \chi \( \dot E^{ij} E_{i}{}^k \hat{S}_{jk}\)  -  H_{S}^B \chi \( \dot B^{ij} B_{i}{}^k \hat{S}_{jk}\)  \Bigg] \,, \\
        \dot J & =- m(Gm)^4 \Bigg[ -2 H_{S}^E \chi \left(E^{ij} E_{ij}\right) + 3 H_{S}^E \chi  \left(E_i{ }^k E_{j k} \hat{s}^i \hat{s}^j \right) + (Gm) H_{\omega}^{E} (\dot E^{ij} E_{i}{}^k \hat{S}_{j k}) \\
& \quad -2 H_{S}^B \chi \left(B^{ij} B_{ij} \right) + 3 H_{S}^B \chi  \left(B_i{ }^k B_{j k} \hat{s}^i \hat{s}^j\right) + (Gm) H_{\omega}^{B} (\dot{B}^{ij} B_{i}{}^k \hat{S}_{j k}) \Bigg] ~\,.
    \end{aligned}
\end{equation}
where the $J$ is the angular momentum. These
equations provide input for 
our waveform
calculations.

We note that one should consider the leading dissipation numbers $H_{\omega}^{E}$ and $H_{\omega}^{B}$ separately. However, for black holes these two parameters turns out to be the same due to the electric-magnetic duality of linear black hole perturbations, which holds for both 
static and spinning black holes in four dimensions~\cite{teukolsky1973perturbations,teukolsky1974perturbations,Chandrasekhar:1985kt,Goldberger:2005cd,Porto:2007qi, Chia:2020yla, Hui:2020xxx}.
In what follows we will 
consider both the general situation, 
and the case when our compact 
black-hole like objects obey the electric-magnetic duality. 

Finally, let us note that  \S\ref{sec:gw_observables}, we will show that, for aligned-spin quasi-circular orbits, the spin-linear $H_{S}^{E}$ and spin-independent $H_{\omega}^{E}$ dissipation parameters first appear in gravitational waveforms at 2.5PN and 4PN respectively. 

\subsection{Estimates for Tidal Dissipation Coefficients}

At the microscopic level, tidal dissipation
of fluid bodies 
is generated by viscosity
\cite{poisson_will_2014,Endlich:2015mke,Ripley:2023qxo}, suggesting:
\begin{equation}
    H_{\omega}^{E/B} \sim \Lambda^{E/B} \times \frac{\tau_d}{Gm} ~, \quad \Lambda^{E/B} \sim  \( \frac{R}{Gm}\)^5 ~,
\end{equation}
where $\tau_d$ is the tidal lag time. In the Eulerian 
fluid dynamics the tidal lag time 
can be expressed through 
the kinematic viscosity $\nu$ and the star radius $R$,
\begin{equation}
    \tau_d \sim \frac{ \nu R}{G m} \,.
\end{equation}
The kinematic viscosity for a general fluid 
can be estimated as $\nu\sim l_{\rm mfp} \langle v\rangle$, where $ l_{\rm mfp}$ is the mean free path
and $\langle v\rangle \sim (k_BT/\mathcal{M})^{1/2}$ is the average velocity of 
fluid particles ($k_B$ is the Boltzmann constant, $T$ is fluid temperature, and $\mathcal{M}$ is the particle mass). For black holes, the dissipation
coefficient is given by 
\be 
    \tau_d \sim \frac{c r^2_s}{G m} \,.
\ee 
If we naively interpret this 
result as coming from fluid dynamics, we would 
conclude that the fluid has a mean free path 
equal to that of the size of the 
compact object and the mean velocity
is equal to $c$. Obviously, this represents
a very extreme scenario that is unlikely to realize 
in any actual microscopic model. This, however, sets a good benchmark
for typical dissipation
coefficients of celestial bodies. Denoting 
$\nu_{\rm BH}\equiv c r_s$, we get
\be 
    H_{\omega}^{E/B} \sim \(\frac{R}{ G m}\)^6 \times \( \frac{\nu}{\nu_{\rm BH}}\) ~, 
\ee 
For usual starts, the mean free path of the ionized gas is $l_{\rm mfp}\sim \frac{k^2_B T^2}{\pi e^4n}\sim 10^{-3}$cm, which gives $\nu\sim 10^3$ cm$^2$sec$^{-1}$, resulting in the following estimate:
\be 
\text{Usual stars:}\quad 
\nu/\nu_{\rm BH}\sim 10^{-13}~\,,
\ee 
where we used $T=10^4$K, $n\sim 10^{16}$cm$^{-3}$, $m=1m_{\odot}$. For neutron stars, the highest realistic 
value of 
the bulk viscosity is 
$\zeta \sim 10^{30}$ g/cm/s~\cite{Most:2021zvc,Most:2022yhe,Alford:2022ufz,Yang:2023ogo}.
For typical densities of $10^{15}$ g/cm$^3$ this implies
\be 
\text{Neutron stars:}\quad 
\nu/\nu_{\rm BH}\sim 0.1~\,,
\ee
where we assumed $m=1.5~m_{\odot}$. Owing to the 
extra enhancement by $(R/r_s)^6\sim 5^6\simeq 10^4$,
we see that the dissipation 
coefficient of neutron stars may be 
non-negligible.

The dissipation parameters of Kerr black holes can be read off from black hole perturbation theory calculations~\cite{Goldberger:2005cd, Chia:2020yla, LeTiec:2020spy, Saketh:2022xjb, future2}:
\be 
\label{eqn:bh_dissipation_general}
        H_{\omega}^{E/B}  = \frac{8}{45}\left(1+ \sqrt{1 - \chi^2} \right) ~,\quad H_{S}^{E/B} = - \frac{8}{45} (1 + 3 \chi^2) \,.
\ee
In the observationally relevant small spin limit we have:
\be 
\label{eq:BH low spin value}
        H_{\omega}^{E/B} (\chi \rightarrow 0) =  \frac{16}{45} ~, \quad H_{S}^{E/B} (\chi \rightarrow 0) = - \frac{8}{45} \,.
\ee
The relationship between $H_{S}^{E/B}$ and $H_{ \omega}^{E/B}$ follows from
superradiance. 
This relationship
is more complext 
for a general body,
where $H^{E/B}_{S}$ depend
on the body's angular velocity
(measured at infinity), implying
\be 
\label{eq:superrad_gen}
H_{S}^{E/B} = - 2 \frac{Gm\Omega}{\chi} H_{\omega} \,.
\ee 
Since 
the angular velocity in general 
is not determined by 
spin~\cite{Cook:1993qr}, 
we could not deduce $H_{S}^{E/B}$
entirely from $H_{\omega}$.
For Kerr black holes, however, 
one can use relations $\Omega^{ab}=\Omega \hat S^{ab}$,  $\Omega = a/(r_+^2+a^2)$, implying
\be 
\label{eq:kerr superradiance relation}
    H_{S}^{E/B}|_{\rm Kerr} = - \frac{1}{2} H_{\omega}^{E/B}|_{\rm Kerr} \,,
\ee
which we have already seen
in Eq.~\eqref{eq:BH low spin value}.
Simulations of rotating quark stars 
in~\cite{Kruger:2019zuz,Chen:2023bxx}
suggest that 
\be 
\Omega \approx \text{const}\times \frac{J}{Gm^2}\left(\frac{Gm}{R^3}\right)^{1/2}\,,\quad \Rightarrow \quad \frac{H_{S}}{H_{\omega}}\sim \left(\frac{r_s}{R}\right)^{3/2}\lesssim 1\,,
\ee 
i.e. 
the dissipation coefficients
associated with angular velocity
should be somewhat suppressed w.r.t. the spin-independent dissipation numbers $H_{\omega}$.



\subsection{Tidal Heating Imprints on Waveforms} \label{sec:gw_observables}

In Fourier space, the gravitational waveform for a quasi-circular aligned-spin binary system in the $(2,2)$ radiation mode takes the general form
\begin{equation}
    \tilde{h}(f) = A(f) \hskip 1pt e^{- i \psi(f)} \, , \quad \tilde{h}_{+}(f) = \tilde{h}(f) \hskip 1pt \frac{1 + \cos^2 \iota}{2} \, , \quad  \tilde{h}_{\times} (f) = - i \tilde{h}(f) \hskip 1pt \cos \iota \label{eqn:freq_amplitude}
\end{equation}
where $A$ is the amplitude, $\psi$ is the phase, $h_{+, \times}$ are the plus and cross GW polarizations, and $\iota$ is the inclination angle between the line of sight and the orbital angular momentum.\footnote{In principle, the functional form of $\iota$ in $h_{+, \times}$ shown in (\ref{eqn:freq_amplitude}) only applies when $A$ contains only the leading order amplitude -- additional angular dependences in $\iota$ would appear when higher order PN corrections to $A$ are taken into account, see e.g.~\cite{VanDenBroeck:2006qu, Arun:2008kb, Mishra:2016whh}. However, in practice we found incorportating PN corrections in the amplitude leads to very little or even no changes to parameter estimation results, and we ignore these detailed angular dependences for simplicity. \label{footnote:Acorrections} } 
We will derive now contributions from tidal dissipation in $\psi$ and $A$ relevant for the inspiral part of the waveform. We will present results for a general waveform that are applicable for any general compact objects,
and a reduced waveform that is useful for 
black hole binary tests.
The latter are simplified by using the electric-magnetic duality and optionally the superradiance relation if we assume the black holes have small spins. 

Let us start with the impact of tidal heating 
on a general compact body. 
We consider the inspiral part of the gravitational waveform, where the imprints on observables can be derived analytically. 
Our derivation here will contain only the 
main results. Details of the intermediate
calculations can be found in Appendix~\ref{appendix:derivation}. 
We start with the energy balance equation
\begin{equation}
   - \mathcal{F}_\infty - \dot M = \dot{\mathcal{E}}  \, , \label{eqn:Ebalance_maintext}
\end{equation}
where $\mathcal{E}$ is the binding energy, $\mathcal{F}_\infty$ is the energy flux radiated to infinity, $M = m_1 + m_2$ is the total mass, $\dot{M} = \dot{m_1} + \dot{m_2}$ is the total tidal heating flux absorbed by both binary components, and the overdot stands for a time derivative. For an aligned spin quasi-circular orbit, the energy flux absorbed by the component mass $m_1$ is (see Appendix~\ref{appendix:derivation})

\begin{equation}
\begin{aligned}
\dot{m_1}(v) & = \left( \frac{{9 H_{1S}^{E} m_1^3 \eta^2 \chi_1}}{{M^3}} \right) v^{15} + \Bigg( -\frac{{9 H_{1S}^{E} m_1^3 (-2 M - m_1 + 3 M \eta) \eta^2 \chi_1}}{{M^4}} \\
& + \frac{{9 H_{1S}^{B} m_1^3 \eta^2 \chi_1}}{{M^3}} \Bigg) v^{17} + \left( \frac{{18 H_{1\omega} m_1^4 \eta^2}}{{M^4}} \right) v^{18} ~,
\end{aligned} \label{eqn: mdot_single_circle_maintext}
\end{equation}
where $v = (\pi M f)^{1/3}$ is the PN velocity and $\eta \equiv m_1 m_2 / M^2$ is the symmetric mass ratio. The expression for $\dot{m_2}$ is obtained by interchanging the $1 \leftrightarrow 2$ indices in Eq.~\eqref{eqn: mdot_single_circle_maintext}. 
Compared to the leading order Einstein quadrupolar flux, $\mathcal{F}^{\rm LO}_\infty = 32 \eta^2 v^{10}/5$, Eq.~\eqref{eqn: mdot_single_circle_maintext} shows that the leading tidal heating effect contributes at 2.5PN order.

\vskip 4pt

In the stationary phase approximation~\cite{Thorne1980Lectures}, the phase can be derived by iteratively solving for the orbital phase $\phi(v)$ and time $t(v)$, both of which depend on $\dot{v}$~\cite{Damour:2000zb, Buonanno:2009zt}. The $\dot{v}$ term can be obtained by applying the chain rule on $\dot{\mathcal{E}}$
\begin{equation}
   \dot{\mathcal{E}} = \frac{\partial \mathcal{E}}{\partial v} \dot{v} + \frac{\partial \mathcal{E}}{\partial m_1} \dot{m_1} + \frac{\partial \mathcal{E}}{\partial m_2} \dot{m_2} ~, \label{eqn:Ebalance_chainrule_maintext}
\end{equation}
where we extended the chain rule to the component masses as well.\footnote{The chain rule could in principle be extended to other instrinsic variables, for instance the component spins which leads to additional terms such as $\dot{\mathcal{E}} \supset \left( \partial  \mathcal{E} / \partial \chi_i \right) \dot{\chi_i} , i = 1,2$. However, these terms appear beyond 3.5PN order in the phase and are therefore neglected in this work.}  
The terms which depend on $\dot{m}_{1,2}$ in (\ref{eqn:Ebalance_chainrule_maintext}) are 
sometimes referred to as ``secular change in mass" and provide additional tidal heating contributions to the phase beyond the component flux (\ref{eqn: mdot_single_circle_maintext}). In particular, since $\partial \mathcal{E} / \partial m \propto v^2$, these terms start contributing to the phase at 3.5PN order~\cite{Isoyama:2017tbp}. The resulting TaylorF2 phase reads
\begin{equation}
    \psi(v) = 2\pi f t_0 - \phi_0 - \frac{\pi}{4} + \frac{3}{128\eta v^5} \Big[  \psi^{\rm pp}(v) + \psi^{\rm TDN} (v)  \Big] ~, \label{eqn:TaylorF2Phase}
\end{equation}
where $t_0$ and $\phi_0$ are the time and phase of coalescence, $\psi^{\rm pp}(v) = 1 + \cdots$ is the point particle contribution which we retain up to 4PN order for the non-spinning terms and up to 3.5PN order for the aligned-spin terms~\cite{Blanchet:2023bwj, Cho:2022syn}; we list these coefficients in (\ref{eqn:phase_PP_general}). 
The phase contributions from the tidal dissipation numbers are (see Appendix~\ref{appendix:derivation})
\be 
\label{eq: dissipation phase}
    \psi^{\rm TDN}(v) = \underbrace{v^5 (1 + 3 \ln v) \hskip 1pt \psi_{\rm 2.5PN}^{\rm TDN}}_{{\text{linear spin}}} \quad + \underbrace{v^7 \hskip 1pt  \psi_{\rm 3.5PN}^{\rm TDN}}_{{\text{linear}} \hskip 1pt +  \hskip 1pt \cdots} + \quad \underbrace{v^8(1 - 3\ln v ) \hskip 1pt \psi_{\rm 4PN}^{\rm TDN}}_{{\text{spin independent}} \hskip 1pt + \hskip 1pt \cdots } ~,
\ee
where the spin dependent terms are retained up to 3.5PN order while the spin-independent phase starts appearing at 4PN order. Note that here we
focus only on the contributions up to linear order in spin; we drop the spin-quadratic, spin cubic and spin quartic terms. 
In this approximation we have:
\be 
    \begin{aligned}
\label{eqn:dissipation_phase_PN}
    \psi_{\rm 2.5PN}^{\rm TDN} & = {\( \frac{25}{4} \mathcal{H}_1^E \chi_s + \frac{25}{4}\overline{\mathcal{H}}_1^E \chi_a \)} ~, \\[2pt]
    \psi_{\rm 3.5PN}^{\rm TDN} & = { \left(\frac{225 }{8} \mathcal{H}_1^B + \frac{102975}{448}  \mathcal{H}_1^E + \frac{675 }{32} \overline{\mathcal{H}}_1^E \delta + \frac{1425 }{16} \mathcal{H}_1^E \eta\right) \chi_s}   \\
        & \quad {+ \left(\frac{225 }{8} \overline{\mathcal{H}}_1^B + \frac{102975 }{448}\overline{\mathcal{H}}_1^E + \frac{675 }{32} \mathcal{H}_1^E \delta+ \frac{1425 }{16}\overline{\mathcal{H}}_1^E \eta \right) \chi_a} \\ 
    \psi_{\rm 4PN}^{\rm TDN} & = {\frac{25}{2} \mathcal{H}^E_0} + \text{other spin-dependent terms}\,,
\end{aligned}
\ee
where $\delta=\left(m_1-m_2\right) / M$ and $\eta=m_1 m_2 / M^2$.
In (\ref{eqn:dissipation_phase_PN}) we introduced several effective mass-weighted tidal dissipation numbers
\begin{equation}
    \begin{aligned}
{ \mathcal{H}^{E/B}_1 \equiv \frac{1}{M^3} \left( m_1^3 H^{E/B}_{1S} + m_2^3 H^{E/B}_{2 S} \right)} ~, & \quad { \overline{\mathcal{H}}^{E/B}_1 \equiv \frac{1}{M^3} \left( m_1^3 H^{E/B}_{1S} - m_2^3 H^{E/B}_{2S} \right)} ~, \\[2pt]
  { \mathcal{H}^{E/B}_0 \equiv \frac{1}{M^4}} & { \left( m_1^4 H^{E/B}_{1 \omega} + m_2^4 H^{E/B}_{2 \omega} \right) } ~,
\end{aligned} \label{eqn:all_effective_dissipation_numbers}
\end{equation}
 which are better measured than the individual component dissipation numbers in data. Note that the spin-dependent terms consist of both symmetric and antisymmetric counterparts, which are labeled without and with an overline respectively. The subscripts $1$ and $0$ denote the linear-in-spin dependence, cubic-in-spin dependence and spin-independence of those coefficients respectively. 
For the spin-independent cases, the component dissipation numbers would combine to form a single effective dissipation number (this is analogous to the leading spin-independent effective Love number, which is often denoted as $\tilde{\Lambda}$ in the literature~\cite{Flanagan:2007ix}). Substituting the black hole values (\ref{eqn:bh_dissipation_general}) into (\ref{eqn:dissipation_phase_PN}), we are able to reproduce known results for the BBH waveform phase, which is known up to 3.5PN order~\cite{Isoyama:2017tbp}. 

We summarize the PN orders at which the effective dissipation numbers affect waveforms for general orbits and for aligned-spin quasi-circular orbits
in Table~\ref{tab:pn_counting}. This table illustrates several interesting patterns in the PN counting that are worth highlighting. Firstly, due to the relative velocity suppression between the magnetic and electric tidal fields $B_{\ell m_\ell} \sim v E_{\ell  m_\ell}$, Eq.~\eqref{eq:response} dictates that all magnetic tidal dissipation effects first appear at 1PN order higher than their electric counterparts. 
Furthermore, due to the fact that $M \omega \sim v^3$, the spin-independent terms $\mathcal{H}^{E/B}_{0}$ are 1.5PN suppressed compared to the linear-spin counterparts $\mathcal{H}^{E/B}_{1}, \overline{\mathcal{H}}^{E/B}_{1}$.

\bgroup
\begin{table}[t!]
    \centering
    \setlength\tabcolsep{4pt}
    {\tabulinesep=1.0mm
    \begin{tabular}{|c|c|c|}
        \hline
         \multirow{2}{*}{}Effective &  PN Orders for   & PN Orders for Aligned-Spin    \\[-2pt]
         Dissipation Numbers & General Orbits & Quasi-Circular Orbits  \\[0pt]
         \hline
\rowcolor{gray!20}{$\mathcal{H}^{E}_{1}, \overline{\mathcal{H}}^{E}_{1}$} & 2.5PN, 3.5PN, $\cdots$ & {2.5PN, 3.5PN, $\cdots$}  \\[0pt]
{$\mathcal{H}^{B}_{1}, \overline{\mathcal{H}}^{B}_{1}$} & 3.5PN, $\cdots$ &{3.5PN, $\cdots$}   \\[0pt]
\rowcolor{gray!20} 
{$\mathcal{H}^{E}_{0}$} & 4PN, $\cdots$ & {4PN, $\cdots$}   \\[3pt]
{$\mathcal{H}^{B}_{0}$} & 5PN, $\cdots$ & {5PN, $\cdots$}   \\[0pt]
         \hline
     \end{tabular}}
    \caption{Summary of the PN orders at which the effective dissipation numbers (\ref{eqn:all_effective_dissipation_numbers}) appear in waveform observables, for both general orbits and for aligned-spin quasi-circular orbits. In this work, we retain the spin-dependent terms up to 3.5PN order and the spin-independent term up to 4PN order. 
    }
    \label{tab:pn_counting}
\end{table}
\egroup

\vskip 4pt

It is important to keep in mind that $\mathcal{H}^{E/B}_1$ are negative definite due to  superradiance. Physically, this negative sign can be interpreted as the energy extraction process increasing the binding energy of the binary, which widens the orbit and leads to a smaller orbital velocity. On the other hand, $\mathcal{H}^{E/B}_0$ are positive definite because they represent energy absorption. 
Finally, the asymmetric effective dissipation numbers $\overline{\mathcal{H}}^{E/B}_1$ can be either positive or negative, depending on the relative sizes of the mass weighting and the dissipation numbers of the binary components.

\vskip 4pt

For completeness, we also computed the imprints of tidal dissipation on the GW amplitude. Similar to the phase (\ref{eqn:TaylorF2Phase}), the total amplitude can be separated into the point-particle terms and the tidal heating terms
\begin{equation}
\begin{aligned}
    A(f) & = \sqrt{\frac{5}{24}} \frac{\mathcal{M}^{5 / 6}}{D\pi^{2/3}} f^{-7 / 6} \Big[ A^{\rm pp}(f) + A^{\rm TDN}(f) \Big] ~,
\end{aligned}
\end{equation}
where $\mathcal{M}$ is the chirp mass and $D$ is the luminosity distance. The point particle terms are shown in (\ref{eqn:amplitude_PP_general}) while tidal dissipation contributes in the following manner
\begin{equation}
    A^{\rm TDN}(f) =   v^5 A_{\rm 2.5PN}^{\rm TDN} + v^7 A_{\rm 3.5PN}^{\rm TDN} + v^8 A_{\rm 4PN}^{\rm TDN} ~, \label{eqn:dissipation_amplitudes}
\end{equation}
with the PN coefficients
\begin{equation}
    \begin{aligned}
    A_{\rm 2.5PN}^{\rm TDN} & = - \frac{45}{64} \mathcal{H}_1^E \chi_s - \frac{45}{64} \overline{\mathcal{H}}_1^E \chi_a ~, \\
        A_{\rm 3.5PN}^{\rm TDN} & = -\left(\frac{45 }{64} \overline{\mathcal{H}}_1^B + \frac{73755 }{14336}\overline{\mathcal{H}}_1^E + \frac{45  }{128} \mathcal{H}_1^E \delta + \frac{465 }{512}  \overline{\mathcal{H}}_1^E \eta \right) \chi_a \\
        & \quad + \left(-\frac{45 }{64} \mathcal{H}_1^B - \frac{73755 }{14336}\mathcal{H}_1^E - \frac{45 }{128} \overline{\mathcal{H}}_1^E \delta - \frac{465 }{512} \mathcal{H}_1^E \eta \right) \chi_s ~, \\
         A_{\rm 4PN}^{\rm TDN} & = - \frac{45}{32} \mathcal{H}^E_0 ~.
    \end{aligned} \label{eqn:dissipation_amplitudes_PN}
\end{equation}
In practice, we find that incorporating these corrections to the waveform amplitude yield marginal to no changes to the parameter estimation results for the O1-O3 events (see also Footnote~\ref{footnote:Acorrections}).

\subsection{Simplified Waveforms for Binary Black Holes}

So far, we have worked without assumptions on the nature of the compact object. However, several simplifications to the waveform can be made if we restrict ourselves to binary black holes  (BBH) and intend to constrain the dissipation parameters of black holes (\ref{eqn:bh_dissipation_general}).
To this end, we can exploit the electric-magnetic duality to reduce number 
of parameters in Eq.~\eqref{eqn:dissipation_phase_PN}. Setting the electric and magnetic dissipation
coefficients to the same values, $\mathcal{H}^B = \mathcal{H}^E$, $\overline{\mathcal{H}}^B = \overline{\mathcal{H}}^E$, we reduce the set of seven tidal dissipation numbers in (\ref{eqn:dissipation_phase_PN}) to five parameters:
\begin{equation}
\begin{aligned}
    \text{E/B Duality for black holes:} & \quad \mathcal{H}^B = \mathcal{H}^E, \hskip 4pt \overline{\mathcal{H}}^B = \overline{\mathcal{H}}^E \\[4pt] \{\mathcal{H}_1^E, \overline{\mathcal{H}}_1^E, \mathcal{H}_1^B, \overline{\mathcal{H}}_1^B, \mathcal{H}_0^E \} & \hskip 2pt \to \hskip 2pt  \{\mathcal{H}_1, \overline{\mathcal{H}}_1, \mathcal{H}_0 \} \, . \label{eqn:EB_duality_constants}
\end{aligned}
\end{equation}
In this case we drop the $E/B$ superscipts for brevity. For future convenience, we will duplicate the tidal dissipation phases below but with the 3.5PN linear-spin term now simplified:
\begin{equation}
    \begin{aligned}
    \psi_{\rm 2.5PN}^{\rm TDN} & = \textcolor{black}{\( \frac{25}{4} \mathcal{H}_1 \chi_s + \frac{25}{4}\overline{\mathcal{H}}_1 \chi_a \)} ~, \\[2pt]
    \psi_{\rm 3.5PN}^{\rm TDN} & = \textcolor{black}{ \left( \frac{115575}{448}  \mathcal{H}_1 + \frac{675 }{32} \overline{\mathcal{H}}_1 \delta + \frac{1425 }{16} \mathcal{H}_1 \eta\right) \chi_s}   \\
        & \quad \textcolor{black}{ + \left(  \frac{115575}{448} \overline{\mathcal{H}}_1 + \frac{675 }{32} \mathcal{H}_1 \delta+ \frac{1425 }{16} \overline{\mathcal{H}}_1 \eta \right) \chi_a} ~, \\ 
    \psi_{\rm 4PN}^{\rm TDN} & = \textcolor{black}{\frac{25}{2} \mathcal{H}_0} + \text{other spin-dependent terms}~ . \label{eqn:dissipation_phase_PN_EBduality}
\end{aligned}
\end{equation}
One could further use the superradiance  condition for black holes (\ref{eq:kerr superradiance relation}), which relates the linear-in-spin and spin-independent dissipation numbers, such that
\begin{equation}
    \begin{aligned}
 \mathcal{H}_1 \approx  - \frac{1}{2M^3} \left( m_1^3 H_{1\omega} + m_2^3 H_{2\omega} \right)~, & \quad \overline{\mathcal{H}}_1 \approx - \frac{1}{2M^3} \left( m_1^3 H_{1\omega} - m_2^3 H_{2\omega} \right) \,.
\end{aligned} 
\end{equation}
Notice that due to different mass scaling in $\mathcal{H}_1$ and $\mathcal{H}_0$ they are not equal to each other in this approximation,
\begin{equation}
  \text{Small-Spin Black Hole Superradiance:} \quad \{\mathcal{H}_1, \overline{\mathcal{H}}_1, \mathcal{H}_0 \} \hskip 2pt \to \hskip 2pt \{\mathcal{H}_1, \overline{\mathcal{H}}_1\}  \, . \label{eqn:small_spin_superradiance_constants}
\end{equation}
This assumption simplifies parameter estimation
even further, reducing the set of independent tidal dissipation parameters to only two.

\section{Constraints on Tidal Heating from LVK O1-O3 Data}
\label{sec:observations}

In this section, we present the parameter estimation (PE) results for the tidal dissipation parameters. We will separate our discussion into two subsections: one that incorporates the tidal dissipation phase contributions (\ref{eqn:dissipation_phase_PN}) to IMRPhenomD~\cite{Khan:2015jqa}, an inspiral-merger-ringdown waveform model for binary black holes (\S\ref{subsection:BBH_PE}), and another that focuses only on the inpiral regime of the IMRPhenomD model (\S\ref{subsection:BSM_PE}).

\subsection{Inspiral-Merger-Ringdown + Dissipation for Binary Black Holes} \label{subsection:BBH_PE}

In this section, we incorporate the tidal dissipation phase contributions (\ref{eqn:dissipation_phase_PN}) into IMRPhenomD~\cite{Khan:2015jqa} --- a waveform model that describes the inspiral-merger-ringdown of quasi-circular aligned-spins BBHs. By virtue of utilizing the BBH merger and ringdown waveforms in the PE process, we are essentially constraining the dissipation numbers of black holes in this analysis. 

\vskip 4pt

Since the tidal dissipation phases (\ref{eqn:dissipation_phase_PN}) are only valid in the inspiral portion of the binary waveform, we incorporate them in the IMRPhenomD model at low frequencies terminate their contributions at high frequencies when the binary approaches merger.
This is achieved by introducing the taping frequency, $f_{22}^{\rm tape}$, which is related to the peak frequency at which the $(2,2)$ radiation mode has maximum amplitude in the merger regime, $f_{22}^{\rm peak}$, via~\cite{Mehta:2022pcn}
\begin{equation}
\label{eq: tape}
    f_{22}^{\rm tape} = \alpha f_{22}^{\rm peak} ~,
\end{equation}
where the constant $0 < \alpha < 1$ is the parameter. In this work, we choose $\alpha = 0.35$, in line with the analysis conducted by the LVK collaboration in the context of testing theories beyond General Relativity~\cite{LIGOScientific:2020tif,LIGOScientific:2021sio}. As discussed in \cite{Mehta:2022pcn}, this choice strikes an good balance between utilizing an appreciable portion of the SNR in inspiral portion of the BBH signal while reducing the contributions from the merger, for which our analytic tidal dissipation phases are invalid. We then model the total GW phase in~\eqref{eqn:freq_amplitude} as
\begin{equation}
\label{eq:IMR_tidal_waveform}
    \psi(f) = 
    \begin{cases}  
         \psi^{\rm IMRPhenomD}(f) + \psi^{\rm TDN} (f) - \psi^{\rm TDN}(f_{22}^{\rm ref}) ~, \quad ~ f \leq f^{\rm tape}_{22} \\
         \psi^{\rm IMRPhenomD}(f) + \psi^{\rm TDN}(f_{22}^{\rm tape}) - \psi^{\rm TDN}(f_{22}^{\rm ref})~, \quad ~ f > f^{\rm tape}_{22} ~, 
    \end{cases} 
\end{equation}
where the phases are $C^0$ continuous at $f=f^{\rm tape}_{22}$. The reference frequency $f_{22}^{\rm ref}$ is the frequency at which the phase of the (22)-mode in IMRPhenomD waveform vanishes, and the value of $\psi^{\rm TDN}(f_{22}^{\rm ref})$ acts merely as an overall constant phase and does not impact the PE. In contrast to the flexible theory-independent approach adopted in \cite{Mehta:2021fgz}, we do not have to apply a smoothing window near the taping frequency in \eqref{eq:IMR_tidal_waveform} as our waveform model and likelihood evaluations are directly implemented in the frequency domain.

\vskip 4pt

Since our analysis in this section only applies to binary black holes, we will impose the black hole E/B duality (\ref{eqn:EB_duality_constants}) to reduce the number of free parameters in the PE process.  We also assume that the black holes have small spins, $\chi \ll 1$, in order to drop the spin-quadratic, spin-cubic and spin-quartic terms in the EFT decomposition described in \S\ref{sec:single_body}. This simplification has the additional advantage of directly relating the spin-linear $H_{1,2 S}$ and spin-independent $H_{1,2\omega}$ dissipation numbers via the superradiance  constraint (\ref{eq:kerr superradiance relation}). As a result, our parameter space spans over 13 dimensions:
the independent intrinsic parameters are $\{m_1, m_2, \chi_1, \chi_2,  H_{1\omega}, H_{2\omega}\}$, which are described in \S\ref{sec:gw_observables}, and the extrinsic parameters include $\{D , t_c ,\phi_c, \iota, \psi, \alpha, \delta\}$, which are the luminosity distance $D$, the coalescence time and phase $t_c$ and $\phi_c$, the inclination angle $\iota$, the polarization angle $\psi$, and the right ascension and declination, $\alpha$ and $\delta$ respectively.
We run our PE studies with \texttt{cogwheel}~\cite{Roulet:2022kot}, and adopt uniform priors for the detector-frame component masses $m_1, m_2$ and for the spin components $\chi_1, \chi_2$, with the spin priors spanning over the interval $\mathbf{U}[-0.99,0.99]$. For the component dissipation parameters $H_{1\omega}, H_{2 \omega}$, we use a uniform prior over the range $\mathbf{U}[-5 \times 10^3, 5 \times 10^3]$. All extrinsic parameters are marginalized analytically for fast PE~\cite{Roulet:2022kot}.

\vskip 4pt

\subsubsection{Constraints for Individual Events} \label{sec:pe_individual}

We conduct PE on the BBH events reported in the O1, O2, O3a and O3b IAS catalog with detector frame chirp masses $\mathcal{M} < 40 M_{\odot}$, as these low-mass binaries have non-negligible portions of their signal dominated by the inspiral part of the waveform in the LIGO-Virgo observing band. We focus on the effective dissipation numbers $\mathcal{H}_1, \overline{\mathcal{H}}_1$ and $\mathcal{H}_0$  as the phase terms in (\ref{eqn:dissipation_phase_PN}) suggest that these mass-weighted quantities are better constrained than the component dissipation numbers. 

\begin{figure*}[b!]
    \centering
    \includegraphics[width=0.66\textwidth, trim=0 10 0 20]{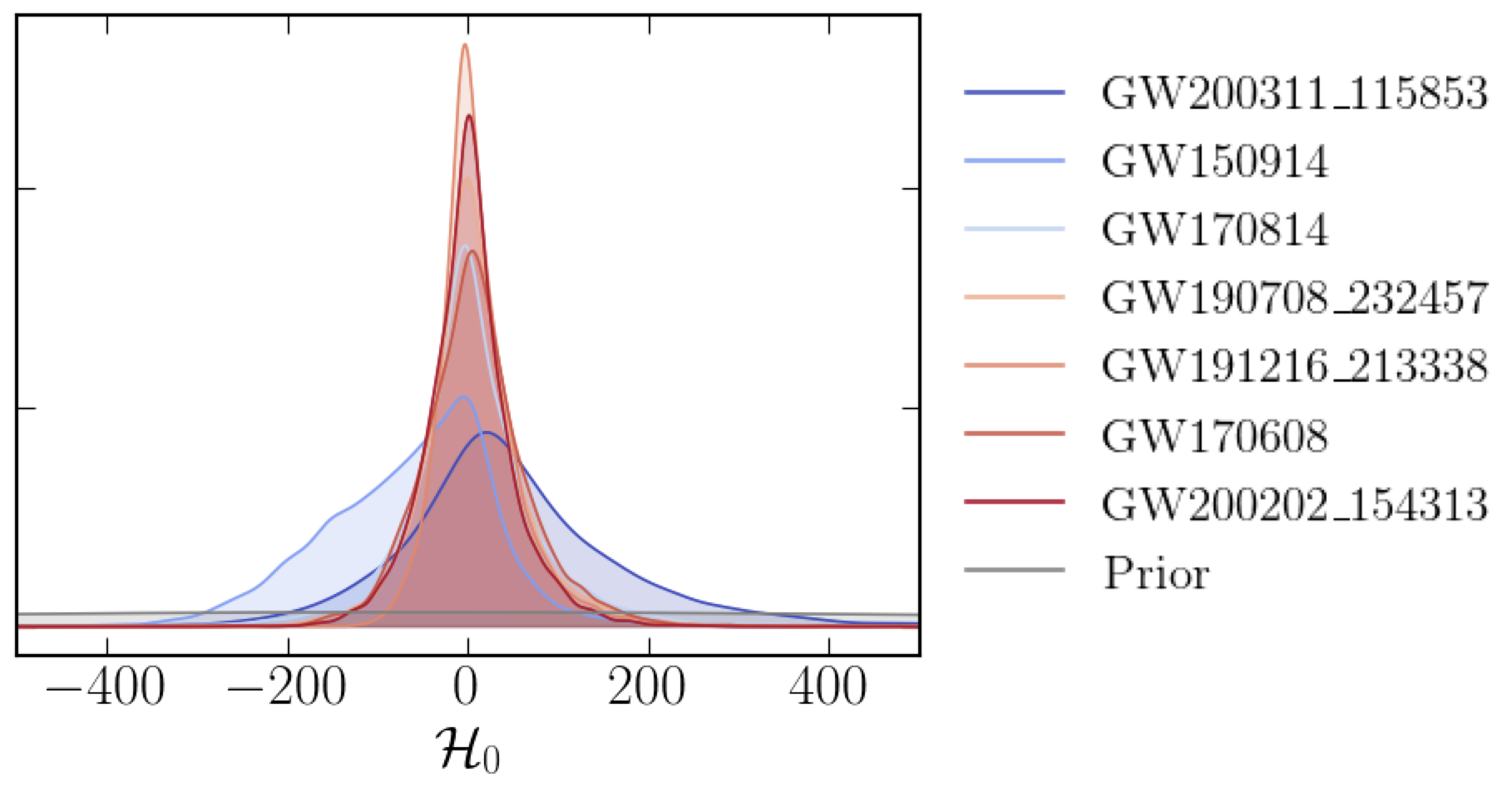}\label{fig:h0} \\
    \begin{adjustbox}{valign=t}
        \includegraphics[width=0.4\textwidth, trim=0 20 0 0]{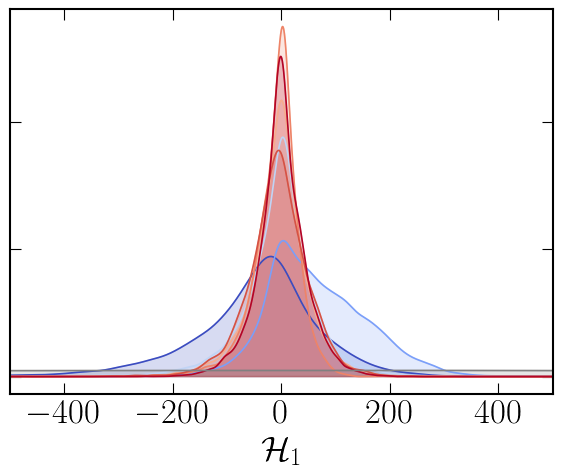}\label{fig:h1}
    \end{adjustbox}
    \begin{adjustbox}{valign=t}
        \includegraphics[width=0.4\textwidth, trim=0 20 0 0]{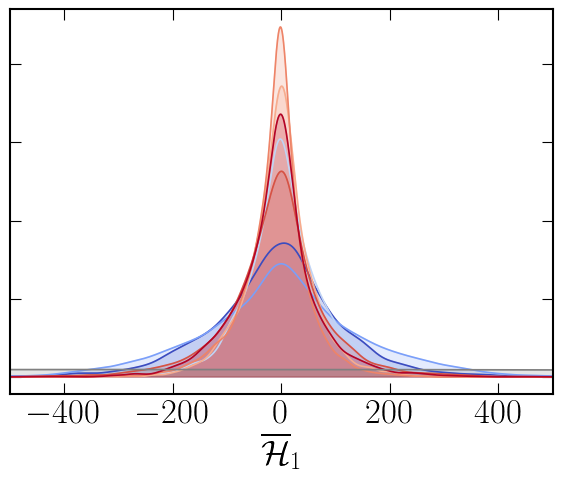}\label{fig:h1bar}
    \end{adjustbox}
    \caption{Marginalized posterior distributions for the spin-independent $\mathcal{H}_0$, mass-symmetric spin-linear $\mathcal{H}_1$, and mass-antisymmetric spin-linear $\overline{\mathcal{H}}_1$ tidal dissipation coefficients for the seven BBH events for which these coefficients are best constrained.}
    \label{fig:H1,H1bar,H0}
\end{figure*}

\begin{figure}[h!]
    \centering
    \includegraphics[scale = 0.34, trim=30 0 0 0]{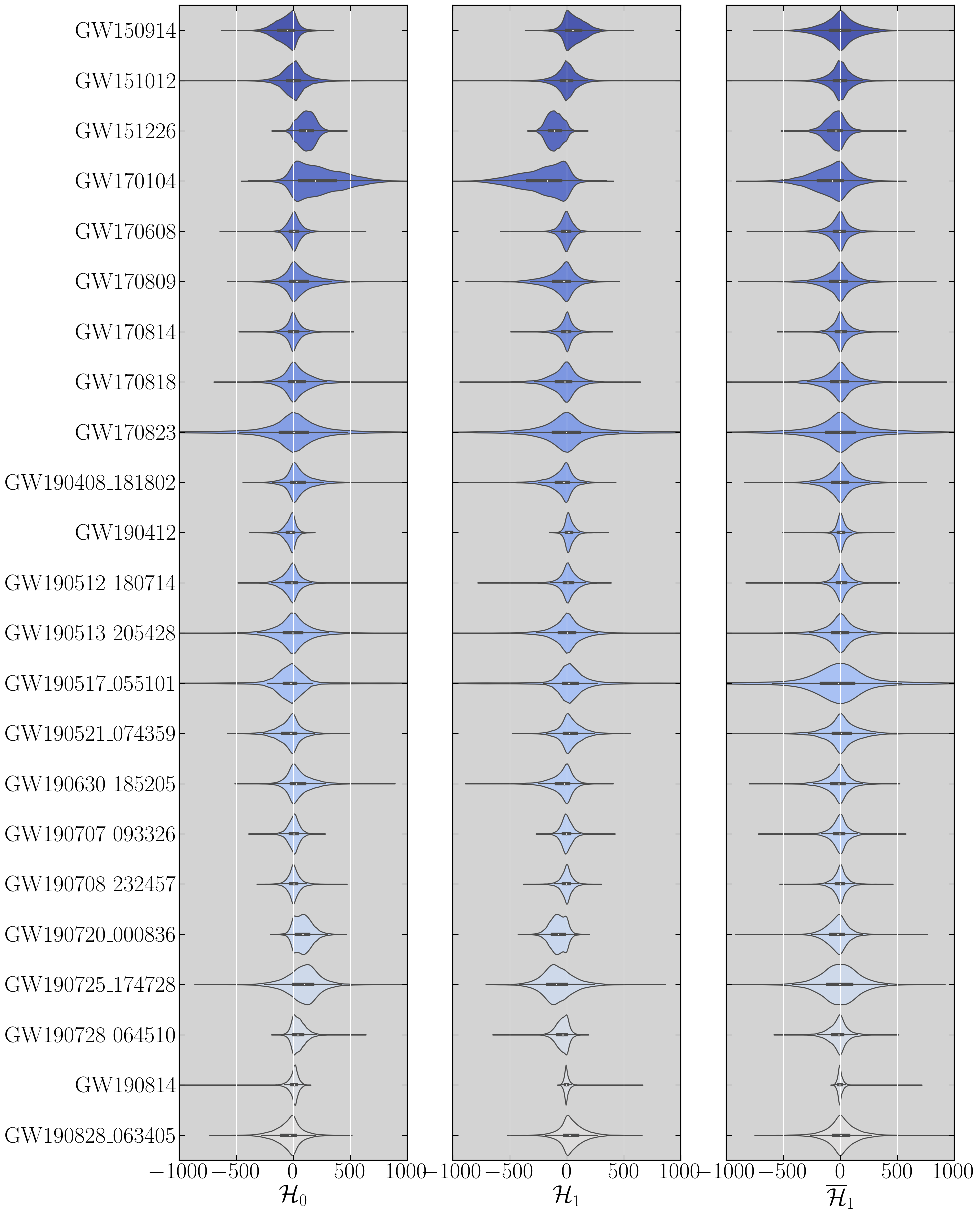}
    \caption{Same as Fig.~\ref{fig:H1,H1bar,H0}, except here we show the marginalized posteriors for the all the BBHs with detector frame chirp mass $\mathcal{M} < 40 M_\odot$ in the IAS O1, O2, O3a and O3b catalog. }
    \label{fig:enter-label}
\end{figure}

\begin{figure}[h!]
    \centering
    \includegraphics[scale = 0.35, trim=30 0 0 0]{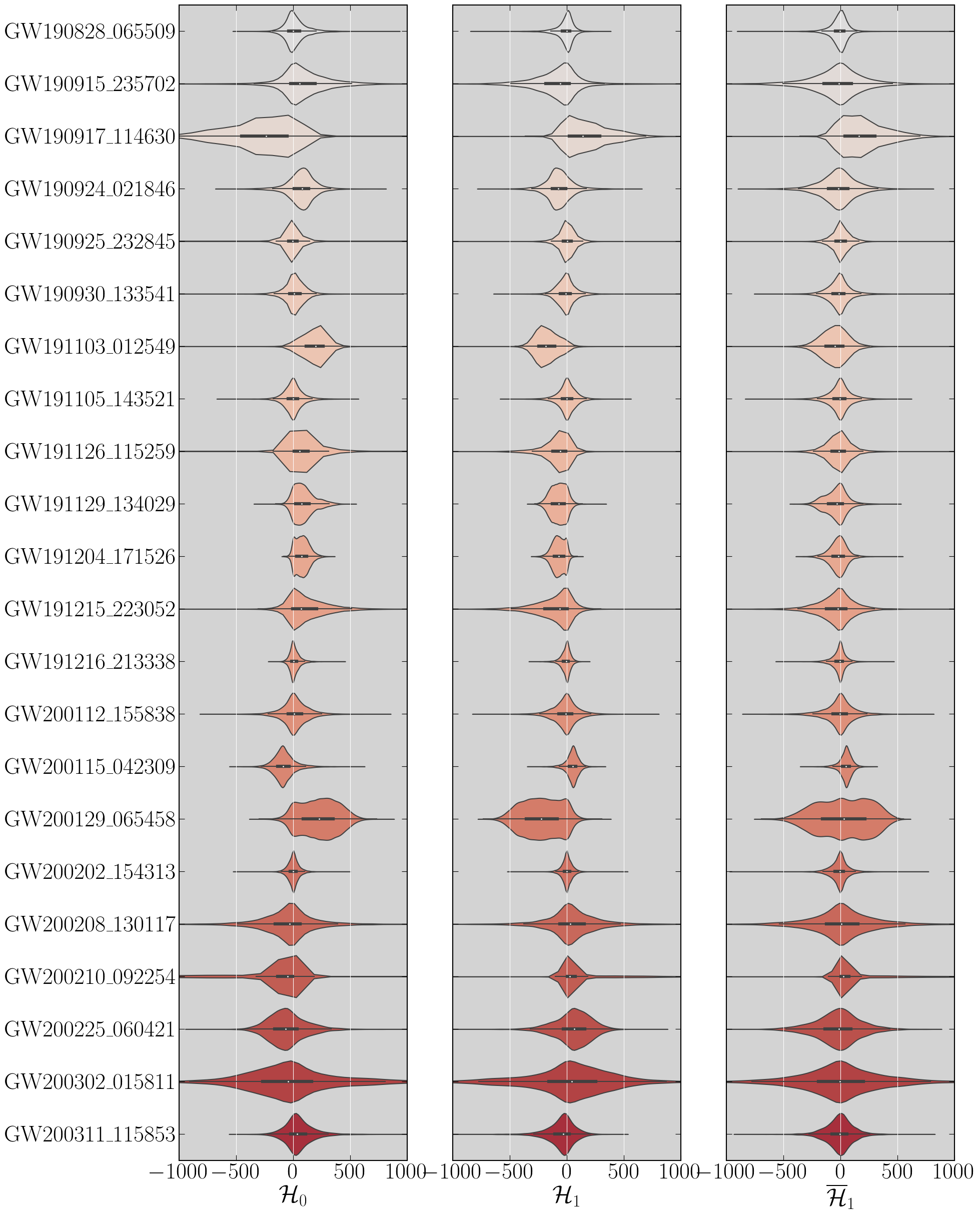}
    \caption{Continuation of Fig.~\ref{fig:enter-label}.}
    \label{fig:enter-label2}
\end{figure}

\vskip 4pt

In Fig.~\ref{fig:H1,H1bar,H0}, we present the marginalized posterior distributions for $\mathcal{H}_1, \overline{\mathcal{H}}_1$ and $\mathcal{H}_0$ for the seven events for which these coefficients are best constrained: GW200311\_115853, GW150914, GW170814, GW190708\_232457, GW191216\_213338, GW170608, GW200202\_154313 (listed in descending order in chirp mass). We observe that these events constrain the dissipation numbers to $|\mathcal{H}_{1}| \sim |\overline{\mathcal{H}}_{1}|\sim |\mathcal{H}_0| \lesssim 300$ at the 90\% credible interval. The constraints tend to be narrower for lower mass binary systems, which is expected as the signal waveforms are dominated by the inspiral portion of the waveform for low mass binaries in the LIGO-Virgo observation bands. For the best event GW191216\_213338, which has a relatively high signal-to-noise ratio of 19.2 and a relatively small median chirp mass $\mathcal{M}_{\rm chirp} = 8.91^{+0.07}_{-0.05} M_{\odot}$ and the error bars indicating the $90\%$ credible interval, the constraints are $|\mathcal{H}_{1}| \sim |\overline{\mathcal{H}}_{1}|\sim |\mathcal{H}_0| \lesssim 100$. In Figs.~\ref{fig:enter-label} and \ref{fig:enter-label2}, we present the constraints for the remaining BBHs in the IAS event catalog. From these posteriors, it is clear that current detector sensitivities lead to likelihoods that tend to rule out large dissipation values $|\mathcal{H}_{1}| \sim |\overline{\mathcal{H}}_{1}|\sim |\mathcal{H}_0| \gtrsim 500$ while the region for smaller values remains prior dominated.

\subsubsection{Constraints at the Population Level} \label{sec:pe_population}

We may obtain a stronger constraint on the black hole dissipation numbers by combining the posterior samples for the individual events shown in Fig.~\ref{fig:enter-label} and \ref{fig:enter-label2}, essentially constraining the dissipation numbers at the level of the black hole population. In what follows, we will also constrain the ratio of energy lost due to tidal dissipation from the total energy flux. For simplicity, we will focus on the spin-independent coefficient $\mathcal{H}_0$ as the results are similar for the other spin-dependent terms.

\begin{figure*}[t!]
    \centering 
   \begin{adjustbox}{valign=t}
        \includegraphics[width=0.47\textwidth, trim=0 10 0 0]{figure/combined_H0.png}
    \end{adjustbox}\\
    \begin{adjustbox}{valign=t}
        \includegraphics[width=0.47\textwidth, trim=0 10 0 0]{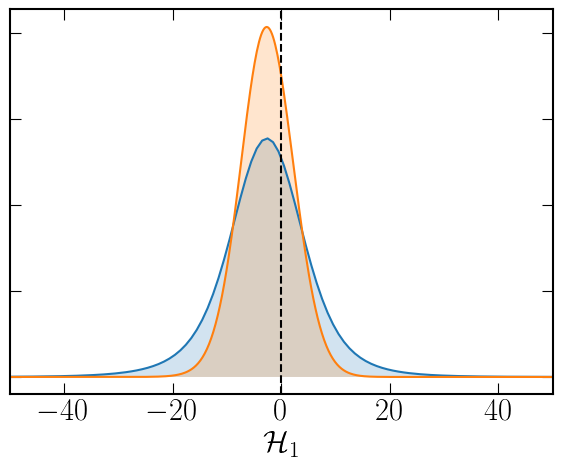}
    \end{adjustbox}
    \begin{adjustbox}{valign=t}
        \includegraphics[width=0.47\textwidth, trim=0 10 0 0]{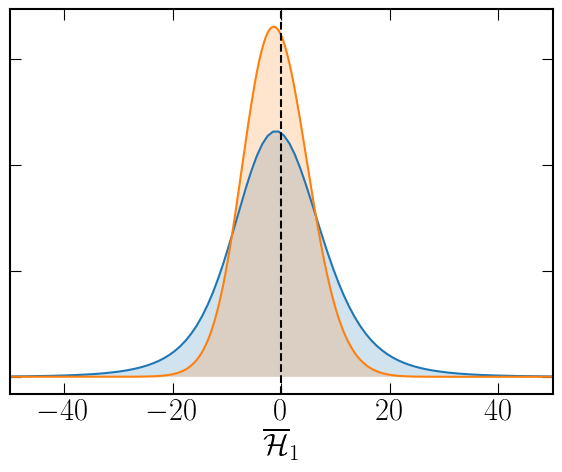}
    \end{adjustbox}
     \caption{Constraints on $\mathcal{H}_0$, $\mathcal{H}_1$ and $\bar{\mathcal{H}}_1$ of the BBH population. We present results for the joint posterior distribution and the posterior computed using the hierarchical Bayesian approach. The black dashed lines represent the theoretical GR prediction for black holes.}
    \label{fig:combined H0,H1,H1bar}
\end{figure*}

We conduct the population inference via two approaches: \textit{i)} by computing the joint posterior distribution by multiplying the posterior samples of the individual events, and \textit{ii)} using the hierarchical Bayesian approach~\cite{Isi:2019asy,Saleem:2021vph}. The joint posterior distribution of the BBH events, assuming they are independent, is proportional to the product of the individual likelihoods
\begin{equation}
    \mathcal{P}(\mathcal{H}_0|\boldsymbol{d}) \propto \mathcal{L}(\boldsymbol{d}|\mathcal{H}_0) \mathcal{P}(\mathcal{H}_0) ~, \quad \mathcal{L}(\boldsymbol{d}|\mathcal{H}_0) = \prod_{j=1}^N \mathcal{L}(d_j|\mathcal{H}_0) ~, \label{eqn:bayes}
\end{equation}
where $\boldsymbol{d} = \{ d_j\}$ is the joint data of the individual events and $\mathcal{P}(\mathcal{H}_0)$ is the prior for $\mathcal{H}_0$. In the first approach, we multiply the marginalized posterior samples $\mathcal{P}(\mathcal{H}_0|d_j)$ collected for the individual BBH events in \S\ref{sec:pe_individual}, and then reweight the final distribution by dividing it with the prior $\mathcal{P}(\mathcal{H}_0)$ of each event (which are all approximately uniform but can vary slightly from one another because $\mathcal{P}(\mathcal{H}_0) $ depends on mass). The result of this approach is shown in Fig.~\ref{fig:combined H0,H1,H1bar}, where the joint constraint is $ -6 < \mathcal{H}_0 <  13 $ at the $90\%$ credible level. For linear-spin dissipation numbers $\mathcal{H}_1$ and $\bar{\mathcal{H}}_1$, we obtain the joint constraints $-11 < \mathcal{H}_1 < 6$ and $-11 < \bar{\mathcal{H}}_1 < 9$ at the $90\%$ credible level respectively. Note that by virtue of the central limit theorem, the joint distributions are Gaussian distributed, even though the posteriors of the individual events in Figs.~\ref{fig:enter-label} and \ref{fig:enter-label2} are generally non-Gaussian.  

In the hierarchical combining approach, we infer the joint posterior $\mathcal{P}(\mathcal{H}_0|\boldsymbol{d})$ by assuming that  $\mathcal{H}_0$ follows an underlying population distribution that is governed by a set of hyperparameters. In this work, we assume that $\mathcal{H}_0$ is Gaussian distributed with mean $\mu$ and standard deviation $\sigma$ at the population level:
\begin{equation}
    \mathcal{P}(\mathcal{H}_0  | \mu,\sigma ) = \mathcal{N}(\mathcal{H}_0 | \mu,\sigma^2) ~ , 
\end{equation}
The joint posterior $\mathcal{P}(\mathcal{H}_0|\boldsymbol{d})$ would be obtained by marginalizing over the posterior distribution on these hyperparameters 
\begin{equation}
    \mathcal{P}(\mathcal{H}_0|\boldsymbol{d}) = \int \mathcal{P}(\mathcal{H}_0|\mu,\sigma) \mathcal{P}(\mu,\sigma|\boldsymbol{d}) d\mu d\sigma ~.
\end{equation}
From Bayes rule, $\mathcal{P}(\mu,\sigma|\boldsymbol{d})$ can be computed as follow:
\begin{equation}
    \mathcal{P}(\mu,\sigma|\boldsymbol{d}) \propto \mathcal{L}(\boldsymbol{d}|\mu,\sigma) \mathcal{P}(\mu,\sigma) ~, \quad \mathcal{P}(\boldsymbol{d}|\mu,\sigma) = \prod_{j=1}^N \mathcal{L}(d_j|\mu,\sigma) ~. \label{eqn:mu_sigma_posterior}
\end{equation}
where the likelihood for the individual events for a given set of $\{\mu,\sigma\}$  is obtained by marginalizing over the dissipation number posterior distribution of the individual events, $\mathcal{H}^j_0$ : 
\begin{equation}
    \mathcal{L}(d_j|\mu,\sigma) = \int \mathcal{L}(d_j|\mathcal{H}^j_0) \mathcal{P}(\mathcal{H}^j_0|\mu,\sigma) d\mathcal{H}^j_0 ~ . \label{eqn:likelihood_dj}
\end{equation}
Strictly speaking, the likelihood of the individual event in the integrand above should be $\mathcal{L}(d_j|\mathcal{H}^j_0, \mu, \sigma)$. However, since $\mathcal{L}(d_j|\mathcal{H}^j_0, \mu, \sigma)$ cannot be directly computed, the conditional dependence on $\{ \mu, \sigma \}$ is dropped -- this is the essence of hierarchical Bayesian inference. We compute $\mathcal{L}(d_j|\mathcal{H}^j_0)$ by dividing the marginalized posteriors in Figs.~\ref{fig:enter-label} and~\ref{fig:enter-label2} by the prior of the individual events $\mathcal{P}(\mathcal{H}^j_0)$ in the PE. Because we assume that the observed distribution on $\mathcal{H}_0^j$ arises from an underlying population, we take $\mathcal{P}(\mathcal{H}^j_0|\mu,\sigma) = \mathcal{P}(\mathcal{H}_0|\mu,\sigma) $ in (\ref{eqn:likelihood_dj}). Finally, to compute (\ref{eqn:mu_sigma_posterior}) we choose the hyperprior $\mathcal{P}(\mu,\sigma) = \mathcal{P}(\mu) \mathcal{P}(\sigma)$ to be independent and uniformly distributed, with $\mu \sim \mathbf{U}[-40,40]$ and $\sigma \sim \mathbf{U}[0, 30]$. The results of the hierarchical approach is also shown in the left panel of Fig.~\ref{fig:combined H0,H1,H1bar}, where we find that the joint constraint is very similar to that from direct multiplications, with {$ -13 < \mathcal{H}_0 <  20 $, $-18 < \mathcal{H}_1 < 11$ and $-18 < \bar{\mathcal{H}}_1 < 16$ at the $90\%$ credible level.

\vskip 4pt

Finally, we use the same two approaches outlined above to constrain the ratio of the energy loss due to tidal dissipation, $\Delta E_{\rm H}$, to the energy loss at infinity, $\Delta E_{\infty}$ (see (\ref{eqn:energy_loss}) in Appendix~\ref{appendix:derivation} for details of the computation). This ratio provides an alternative and more physically interpretable way of understanding the importance of tidal dissipation in the binary dynamics. The results are shown in Fig.~\ref{fig:summary_BBH}, where the constraint of $\Delta E_{\rm H}/\Delta E_{\infty}$ at the population level is $-0.0026<\Delta E_{\rm H}/\Delta E_{\infty} < 0.0025$ at $90\%$ credible level using the hierarchical approach. This finding is consistent with the GR prediction that BBHs would lose approximately a faction of $\sim 10^{-5} - 10^{-4}$ of its energy to tidal dissipation compared to the radiation at infinity.

\subsection{Inspiral-only + Dissipation for Exotic Binaries} \label{subsection:BSM_PE}

Unlike in \S\ref{subsection:BBH_PE}, where we used the full inspiral-merger-ringdown waveforms of IMRPhenomD as a baseline model, in this section we restrict ourselves only to the inspiral regime of IMRPhenomD. This inspiral-only waveform, which is similar to the analytic TaylorF2 waveform model~\cite{Buonanno:2009zt,Arun:2004hn} up to non-perturbative resummations of non-linear effects such as through Padé resummation \cite{Damour:1997ub,Damour:2007yf} and parameterized higher PN terms \cite{Husa:2015iqa,Khan:2015jqa}, applies to the quasi-circular inspiral of all types of binary systems, including binary black holes and exotic binary systems. As a result, we \textit{do not} assume that that the signals in the GW catalog are sourced by BBHs in this section. This inspiral-only analysis allows us to present our results as general constraints that apply to all types of exotic compact binary coalescences.

\begin{figure*}[t!]
    \centering
    \includegraphics[width=0.56\textwidth, trim=0 10 0 20]{figure/ins_H0.png} \\
    \begin{adjustbox}{valign=t}
        \includegraphics[width=0.34\textwidth, trim=0 10 0 0]{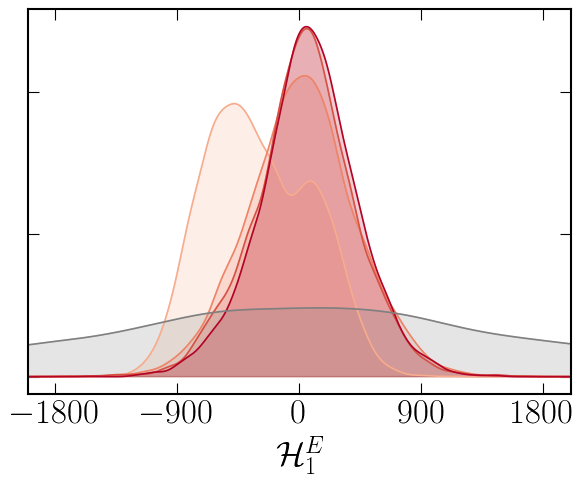}
    \end{adjustbox}
    \begin{adjustbox}{valign=t}
        \includegraphics[width=0.34\textwidth, trim=0 10 0 0]{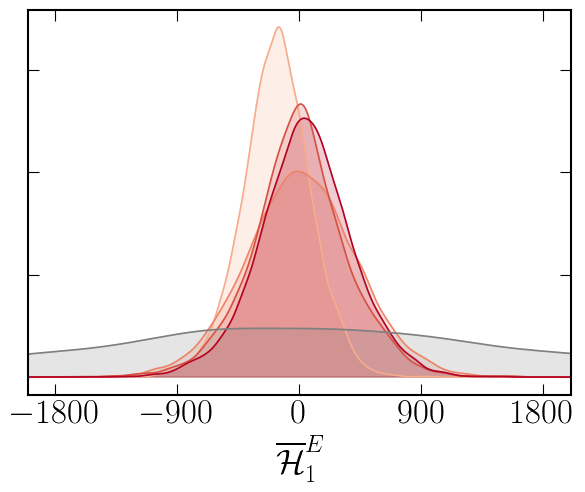}
    \end{adjustbox}\\
    \begin{adjustbox}{valign=t}
        \includegraphics[width=0.34\textwidth, trim=0 10 0 0]{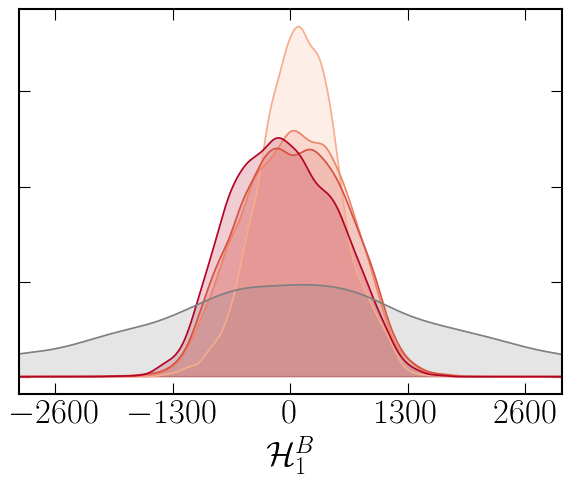}
    \end{adjustbox}
    \begin{adjustbox}{valign=t}
        \includegraphics[width=0.34\textwidth, trim=0 10 0 0]{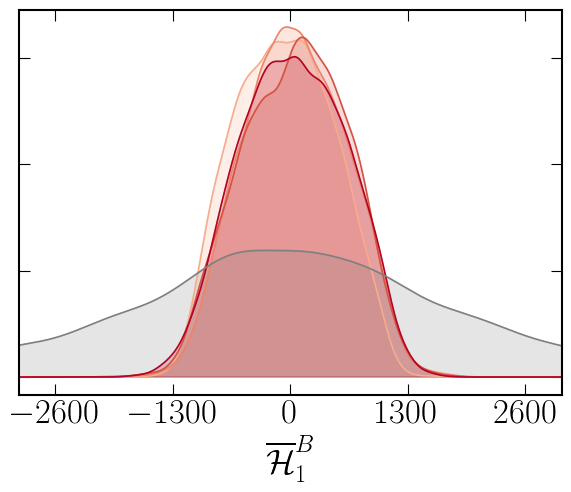}
    \end{adjustbox}\\
    \begin{adjustbox}{valign=t}
        \includegraphics[width=0.34\textwidth, trim=0 10 0 0]{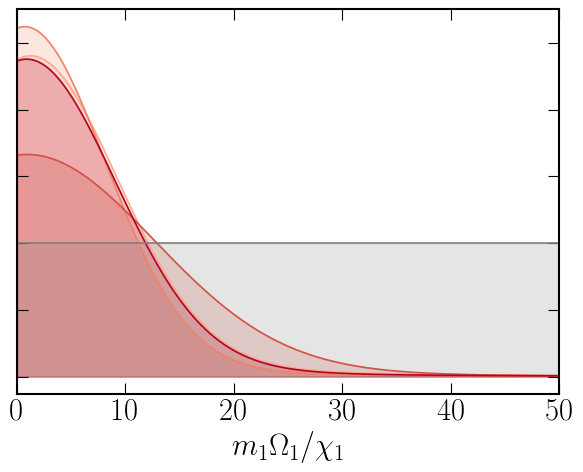}
    \end{adjustbox}
        \begin{adjustbox}{valign=t}
        \includegraphics[width=0.34\textwidth, trim=0 10 0 0]{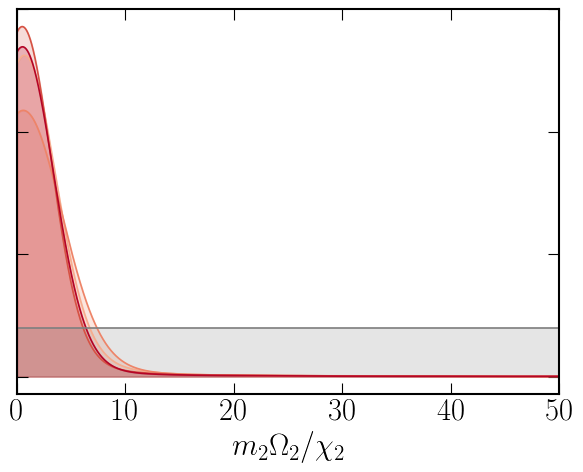}
    \end{adjustbox}
     \caption{Similar to Fig.~\ref{fig:H1,H1bar,H0}, except here we incorporate the tidal dissipation coefficients in a baseline model which consists only of the inspiral regime of IMRPhenomD (we only show the same four events with $\mathcal{M} < 25 M_\odot$). Since the inspiral-only waveform applies not only to binary black holes but also to all types of binary coalescences, we present these results as general constraints for exotic compact objects. Since all astrophysical objects other than black holes do not respect E/B duality, we treat the electric (second row) and magnetic (third row) linear-spin dissipation numbers separately. In the final row, we include the posteriors for dimensionless angular velocity of the two compact objects.}
    \label{fig:ins_H1,H1bar,H0,h3,h3bar}
\end{figure*}

In order to focus on the inspiral regime, we project out the merger-ringdown portions in both the data, $d$, and the IMRPhenomD model, $h$. More precisely, we achieve this projection by truncating both $d$ and $h$ in the frequency domain via the new likelihood
\begin{equation}
    \log \mathcal{L}_{\rm new} = \langle d|\mathbb{P}| h\rangle-\frac{1}{2}\langle h|\mathbb{P}| h\rangle \, ,
\end{equation}
where the projection operator $\mathbb{P}$ is a top hat window function in Fourier space
\begin{equation}
\langle a|\mathbb{P}| b\rangle \equiv 4 \operatorname{Re} \int_{f_{\rm min}}^{f_{\mathrm{22}}^{\rm tape}} \frac{\tilde{a}^*(f) \tilde{b}(f)}{S_n(f)} ~,
\end{equation}
with $f_{\rm min} = 20 $Hz being the minimum frequency and $f_{22}^{\rm tape}$ is the tapering frequency in (\ref{eq: tape}), which determines the frequency range at which the binary inspiral is separated from the merger and ringdown. 

Since we do not assume that the signals arise from black holes, we do not impose the E/B duality in the following PE analysis. We still assume the objects have small spins, $\chi \ll 1$, in order to ignore the spin-quadratic, spin-cubic and spin-quartic dissipation terms. Having said that, the superradiance relation (\ref{eq:superrad_gen}) for a general object in the $\chi \ll 1$ limit would not directly relate the spin-linear and spin-independent terms due to the presence of $\Omega$ as another free parameter.
As a result, our parameter space is now widened to 17 dimensions, where the extrinsic parameters are the same as before and the independent intrinsic parameters are $\{m_1, m_2, \chi_1, \chi_2,  H^B_{1 S}, H^B_{2S}, H_{1 \omega}, H_{2\omega},\Omega_1, \Omega_2  \}$. We use flat prior for the dimensionless angular momentum $m_i \Omega_i / \chi_i \sim \mathbf{U}[0,10^2], i=1,2$.

In Fig.~\ref{fig:ins_H1,H1bar,H0,h3,h3bar}, we show the marginalized posterior distributions of the spin-independent dissipation number $\mathcal{H}_0$, the electric and magnetic, mass-symmetric and antisymmetric linear-spin dissipation numbers, $\mathcal{H}^{E,B}_{1}, \overline{\mathcal{H}}^{E,B}_{1}$. We present the results for the same four lightest events shown in Fig.~\ref{fig:H1,H1bar,H0} and omit the heavier events with $\mathcal{M} > 25 M_\odot$, as the waveforms of these events in the LIGO-Virgo observing bands are dominated by the merger regime where our inspiral-only waveform is invalid. The constraints are now manifestly wider than those in Fig.~\ref{fig:H1,H1bar,H0}; for instance, we find $|\mathcal{H}_0| \lesssim 400$ at $90\%$ credible interval here compared to the  $90\%$ credible bound of $|\mathcal{H}_0| \lesssim 100$ in the inspiral-merger-ringdown analysis in \S\ref{sec:pe_individual}. The broader posteriors here arise due to the reduced SNR in the likelihood evaluation for the inspiral-only waveform. In particular, since the effective dissipation numbers depend on mass, the lack of merger implies the masses are less precisely measured, resulting in broader posterior distributions in the dissipation numbers.

\vskip 4pt

It is also important to emphasize that, without imposing the kinematic superradiance constraint in this section, $\mathcal{H}_0$ is more precisely measured than the linear-spin dissipation numbers even though $\mathcal{H}_0$ formally appears at 4PN in the waveform phase while the linear-spin terms first appear at a lower 2.5PN order. This is because spins are not very precisely measured by current detectors, and most of the inferred spins are consistent with zero. This key observation motivates $\mathcal{H}_0$ as the most well motivated target parameter in studies of tidal dissipation in binary systems. Indeed, $\mathcal{H}_0$ directly plays an analogous role to the effective tidal Love parameter $\Lambda$, which is the leading tidal deformability parameter in GW analysis. 

In addition, notice that the magnetic dissipation numbers are less well-constrained than the electric counterparts, as is expected because the former first appears at a higher order of 3.5PN while the latter first appear at 2.5PN order (see Table~\ref{tab:pn_counting} for summary). Note that we do not assume here that these exotic binary systems arise from the same underlying population, i.e. unlike \S\ref{sec:pe_population} we do not conduct a population study in this section.

\section{Conclusions and Outlook} \label{sec:conclusions}

In this paper, we present the first rigorous derivation of tidal dissipation of a general rotating body and their associated impacts on GW observables in merging binary systems. Our approach utilizes the worldine EFT framework, whereby the spin-dependent and spin-independent tidal dissipation coefficients are inferred as operators that break time-reversal symmetry in the retarded tidal response function. A summary of the linear-spin, cubic-spin and spin-independent effective dissipation numbers that are best measured in binary systems are shown in (\ref{eqn:all_effective_dissipation_numbers}). We also derived the tidal heating imprints on the GW phases and amplitudes for quasi-circular orbits; these results are presented in (\ref{eqn:dissipation_phase_PN}) and (\ref{eqn:dissipation_amplitudes_PN}). These results \textit{a priori} make no assumption about the nature of the binary components and can therefore be used to test all types of binary constituents, including black holes, neutron stars and other types of exotic compact objects. If one assumes that the binary source is a BBH, one may reduce the dimensionality of the parameter space by imposing the black hole electric-magnetic duality (\ref{eqn:EB_duality_constants}). In this case the tidal heating phase imprints simplify slightly, and we list them in (\ref{eqn:dissipation_phase_PN_EBduality}) for convenience.

\vskip 4pt

A summary of our PE constraints are presented in Table~\ref{tab:pn_counting}, with further details of the methods and results elaborated in Section~\ref{sec:observations}. For BBHs, we evaluated the posterior distributions of the dissipation numbers in Figs.~\ref{fig:H1,H1bar,H0}, \ref{fig:enter-label} and \ref{fig:enter-label2} for the BBHs in the O1-O3 catalog. We further combined these posteriors and obtain a joint posterior on $\mathcal{H}_0$, obtaining the BBH population constraint of $-13 < \mathcal{H}_0 < 20$ at the $90\%$ credible level shown in Fig.~\ref{fig:combined H0,H1,H1bar}. While this result is still two orders of magnitude larger than the theoretical value of $\mathcal{H}_0 = 2/45 \approx 4.4 \times 10^{-2}$ for equal mass BBH, future GW observations with improved detector sensitivity and rapidly increasing number of detections will significantly improve upon current constraints. On the other hand, if we do not assume that the binaries are BBHs, the constraints on the electric and magnetic dissipation numbers are relaxed by an approximately an order of magnitude; see Fig.~\ref{fig:ins_H1,H1bar,H0,h3,h3bar} for the posteriors for a select few events. These results can be interpreted as constraints imposed on exotic types of binary systems as the BH electric-magnetic duality are not imposed in the PE computation.

\vskip 4pt

Our work represents the first holistic treatment of tidal heating effects in GW binary sources. In a way, the derivation of the effective dissipation numbers in our work completes the long-standing goal of finding physically-motivated parameterizations of all leading order finite-size effects of compact binary coalescence. Along with the spin-induced multipole moments and the tidal deformability parameters, we are now well-equipped to build GW models which incorporate all finite-size effects of astrophysical bodies as free parameters in the waveforms. Such waveform models would offer a robust and theoretically-consistent framework for probing new types of compact binary systems: both at the search level, such as a recent work dedicated to searching for compact objects with large Love numbers~\cite{Chia:2023tle}, and for accurate parameter estimations to test for new physics~\cite{LIGOScientific:2021sio}. As the LVK detector sensitivity improves over time and we observe more high signal-to-noise ratio events, it would also be interesting to apply our results in this work to further measure or constrain the dissipation numbers of BBHs. We intend to pursue these interesting research directions in future work.

\begin{acknowledgments}
We thank Anna Biggs, Jingping Li, Ajit Kumar Mehta, Julio Parra-Martinez, Irvin Martínez-Rodríguez, Javier Roulet, Muddu Saketh and Matias Zaldarriaga for stimulating discussions. 
HSC gratefully acknowledges support from the Ambrose Monell Foundation and the Sivian Fund at the Institute for Advanced Study. 

\vskip 8pt

This research has made use of data, software and/or web tools obtained from the Gravitational Wave Open Science Center (\href{https://www.gw-openscience.org/}{https://www.gw-openscience.org/}), a service of LIGO Laboratory, the LIGO Scientific Collaboration and the Virgo Collaboration. LIGO Laboratory and Advanced LIGO are funded by the United States National Science Foundation (NSF) as well as the Science and Technology Facilities Council (STFC) of the United Kingdom, the Max-Planck-Society (MPS), and the State of Niedersachsen/Germany for support of the construction of Advanced LIGO and construction and operation of the GEO600 detector. Additional support for Advanced LIGO was provided by the Australian Research Council. Virgo is funded, through the European Gravitational Observatory (EGO), by the French Centre National de Recherche Scientifique (CNRS), the Italian Istituto Nazionale di Fisica Nucleare (INFN) and the Dutch Nikhef, with contributions by institutions from Belgium, Germany, Greece, Hungary, Ireland, Japan, Monaco, Poland, Portugal, Spain.
\end{acknowledgments}

\appendix

\section{Derivation of Waveform Observables} \label{appendix:derivation}

In this appendix, we present the derivation for the TaylorF2 waveform phase and amplitude which includes arbitrary values of dissipation constants --- up to 3.5PN order for spin-dependent dissipation terms and up to 4PN order for the spin-independent terms. The basic inputs in the derivation are the binding energy $\mathcal{E}$, the energy flux at infinity $\mathcal{F}_{\infty}$, and the tidal heating energy fluxes for the binary components $\dot{m_1}$ and $\dot{m_2}$.

\vskip 4pt 

Since our focus is on tidal heating of general rotating bodies, for simplicity we restrict ourselves to the expressions derived for black holes in $\mathcal{E}$ and $\mathcal{F}_{\infty}$. For example, for the spin-spin interaction terms we set the spin-induced moments~\cite{Krishnendu:2017shb} to the black hole value~\cite{Geroch:1970cd, Hansen:1974zz, Thorne:1980ru}. On the other hand, for $\dot{m_1}$ and $\dot{m_2}$ we present the derivations for general rotating bodies. This will allow us to present the tidal dissipation imprints on the phase and amplitude for both general binary systems and for BBHs. 

\vskip 4pt

We restrict ourselves to aligned-spin orbits by using the following spin variables
\begin{equation}
    \begin{aligned}
        S_\ell = \frac{\boldsymbol{\ell} \cdot \boldsymbol{S}}{G M^2} = \frac{1}{4} \[ (1+ \delta)^2 \chi_1 + (1- \delta)^2 \chi_2 \] ~, \\
        \Sigma_\ell = \frac{\boldsymbol{\ell} \cdot \boldsymbol{\Sigma}}{GM^2} = - \frac{1}{2} \[ (1 + \delta) \chi_1 - (1-\delta) \chi_2 \] ~,
    \end{aligned}
\end{equation}
where $\boldsymbol{\ell}$ is the orbital angular momentum normal vector and the spin vectors are
\begin{equation}
    \boldsymbol{S} = \boldsymbol{S}_1 + \boldsymbol{S}_2 ~, \quad \boldsymbol{\Sigma} = M \( \frac{\boldsymbol{S}_2}{m_2} - \frac{\boldsymbol{S}_1}{m_1}\) ~.
\end{equation}
Furthermore, we restrict ourselves to the $(2,2)$ radiation mode. This implies that the orbital velocity, which is the PN perturbation parameter, is $v = (\pi M f)^{1/3}$. Finally, we assume the orbits are quasicircular.

\subsection*{Binding Energy}

We express the binding energy as
\begin{equation}
    \mathcal{E}(v) = - \frac{M \eta v^2}{2} \left[ \mathcal{E}_{\rm NS}(v) + \mathcal{E}_{\rm SO}(v) v^3 + \mathcal{E}_{\rm SS}(v) v^4 + \mathcal{E}_{\rm SSS}(v) v^7 \right] ~, \label{eqn:binding}
\end{equation}
where the non-spinning point-particule contribution $E_{\rm NS}$ is shown up to 4PN~\cite{Blanchet:2023bwj}:
\begin{equation}
    \begin{aligned}
    \mathcal{E}_{\rm NS}(v) & = 1-\left(\frac{3}{4}+\frac{1}{12} \eta \right) v^2-\left(\frac{27}{8}-\frac{19}{8} \eta +\frac{1}{24} \eta^2\right) v^4 \\
    & \quad -\left\{\frac{675}{64}-\left(\frac{34445}{576}-\frac{205}{96} \pi^2\right) \eta+\frac{155}{96} \eta^2+\frac{35}{5184} \eta^3\right\} v^6 \\
    & \quad + \left\{-\frac{3969}{128}+\left[-\frac{123671}{5760}+\frac{9037}{1536} \pi^2+\frac{896}{15} \gamma_{\mathrm{E}}+\frac{896}{15} \ln (4 v)\right] \eta \right. \\
    & \left.\quad +\left[-\frac{498449}{3456}+\frac{3157}{576} \pi^2\right] \eta^2+\frac{301}{1728} \eta^3+\frac{77}{31104} \eta^4\right\} v^8 .
\end{aligned}
\end{equation}
The logarithmic term above arises from the tail effect to the radiation-reaction potential.
The binding energy from spin-orbital coupling is
\begin{equation}
    \begin{aligned}
        \mathcal{E}_{\rm SO} (v) & = \( \frac{14}{3} S_\ell + 2 \delta \Sigma_\ell \) + \[ \(11 - \frac{61}{9} \eta\) S_\ell + \(3 - \frac{10}{3} \eta\) \delta \Sigma_\ell \] v^2 \\
        & \quad + \[ \(\frac{135}{4} - \frac{367}{4} \eta + \frac{29}{12} \eta^2\) S_\ell + \(\frac{27}{4} - 39 \eta + \frac{5}{4} \eta^2\) \delta \Sigma_\ell  \] v^4 ~.
    \end{aligned}
\end{equation}
Taking the spin-induced quadrupole and octupole to be the black hole values~\cite{Geroch:1970cd, Hansen:1974zz, Thorne:1980ru}, the quadratic-in-spin contribution to the binding energy is 
\begin{equation}
\begin{aligned}
    \mathcal{E}_{\rm SS}(v) & = -4 S_{\ell}^2 - 4 \delta S_{\ell} \Sigma_{\ell} + \left(-1 + 4 \eta\right) \Sigma_{\ell}^2 \\
    & \quad + \left[S_{\ell}^2 \left(-\frac{25}{9} + \frac{10 \eta}{3}\right) + S_{\ell} \left(\frac{10 \delta}{3} + \frac{10 \delta \eta}{3}\right) \Sigma_{\ell} + \left(\frac{5}{2} - \frac{15 \eta}{2} - \frac{10 \eta^2}{3}\right) \Sigma_{\ell}^2\right] v^2 ~, 
\end{aligned}
\end{equation}
whereas the cubic-in-spin terms are
\begin{equation}
    \mathcal{E}_{\rm SSS}(v) = -8 S_{\ell}^3 - 16 \delta S_{\ell}^2 \Sigma_{\ell} + \left(-10 + 40 \eta\right) S_\ell \Sigma_{\ell}^2 + \left(-2 \delta + 8 \delta \eta\right) \Sigma_{\ell}^3 ~.
\end{equation}

\vskip 2pt

\subsection*{Energy Flux at Infinity}

The energy flux radiated to infinity is organized as
\begin{equation}
    \mathcal{F}_{\infty}(v) = \frac{32}{5} \eta^2 v^{10} \left[ \mathcal{F}_{\rm NS}(v)+ \mathcal{F}_{\rm SO}(v) v^3 + \mathcal{F}_{\rm SS}(v) v^4 + \mathcal{F}_{\rm SSS}(v) v^7 \right] ~, \label{eqn:flux}
\end{equation}
where $\mathcal{F}_{\rm NS}(v)$ has recently been derived up to 4PN~\cite{Blanchet:2023soy,Trestini:2023ssa}
\begin{equation}
    \begin{aligned}
    \mathcal{F}_{\rm NS}(v) & = 1 + \left(-\frac{1247}{336} - \frac{35}{12} \eta\right) v^2 + 4 \pi v^3 + \left(-\frac{44711}{9072} + \frac{9271}{504} \eta + \frac{65}{18} \eta^2\right) v^4    \\
    & \quad + \Big(-\frac{8191}{672} - \frac{583}{24} \eta \Big) \pi v^5 + \Bigg[\frac{6643739519}{69854400} + \frac{16}{3} \pi^2 - \frac{1712}{105} \gamma_E - \frac{1712}{105} \ln{(4 v)}  \\
    & \quad + \Big(-\frac{134543}{7776}  + \frac{41}{48} \pi^2 \Big) \eta - \frac{94403}{3024} \eta^2 - \frac{775}{324} \eta^3\Bigg] v^6 + \Big(-\frac{16285}{504} + \frac{214745}{1728} \eta  \\
    & \quad + \frac{193385}{3024} \eta^2 \hskip 2pt \Big) \pi v^7  + \Bigg[-\frac{323105549467}{3178375200} + \frac{232597}{4410} \gamma_E - \frac{1369}{126} \pi^2 + \frac{39931}{294} \ln{2} \\
    & \quad - \frac{47385}{1568} \ln{3} + \frac{232597}{4410} \ln{v} + \Big(-\frac{1452202403629}{1466942400} + \frac{41478}{245} \gamma_E - \frac{267127}{4608} \pi^2 \\
    & \quad + \frac{479062}{2205} \ln{2} + \frac{47385}{392} \ln{3}  + \frac{41478}{245} \ln{v} \hskip 2pt \Big) \eta + \left(\frac{1607125}{6804} - \frac{3157}{384} \pi^2\right) \eta^2 \\
    & \quad + \frac{6875}{504} \eta^3 + \frac{5}{6} \eta^4 \Bigg] v^8 ~.
    \end{aligned}
\end{equation}
The logarithmic term which appears at 3PN arises from the tail effect of the renormalization group (RG) running of the quadrupole moment. At 4PN, not only are there the RG contributiosn but also the tail-of-memory effects \cite{Blanchet:2023soy,Trestini:2023ssa}. The spin-orbital coupling contribution to the flux at infinity is
\begin{equation}
    \begin{aligned}
\mathcal{F}_{\rm SO}(v) & = -4 S_{\ell} - \frac{5 \delta \Sigma_{\ell}}{4} + v^2 \left[\left(-\frac{9}{2} + \frac{272 \eta}{9}\right) S_{\ell} + \left(-\frac{13 \delta}{16} + \frac{43 \delta \eta}{4}\right) \Sigma_{\ell}\right] \\
& \quad - 16 \pi v^3 S_{\ell} - \frac{31 \delta \pi \Sigma_{\ell} v^3}{6} \\
& \quad + v^4 \left[\left(\frac{476645}{6804} + \frac{6172 \eta}{189} - \frac{2810 \eta^2}{27}\right) S_{\ell} + \left(\frac{9535 \delta}{336} + \frac{1849 \delta \eta}{126} - \frac{1501 \delta \eta^2}{36}\right) \Sigma_{\ell}\right] ~, 
    \end{aligned}
\end{equation}
Similar to the binding energy, we take the black hole spin-induced multipole moment values~\cite{Geroch:1970cd, Hansen:1974zz, Thorne:1980ru} for the spin-spin contributions to the flux. For quadratic spins, this is
\begin{equation}
    \begin{aligned}
    \mathcal{F}_{\rm SS}(v) & = 8 S_{\ell}^2 + 8 \delta S_{\ell} \Sigma_{\ell} + \left(\frac{33}{16} - 8 \eta\right) \Sigma_{\ell}^2 \\
    & \quad + \left(\left(-\frac{3839}{252} - 43 \eta\right) S_{\ell}^2 + \left(-\frac{1375 \delta}{56} - 43 \delta \eta\right) S_{\ell} \Sigma_{\ell} + \left(-\frac{227}{28} + \frac{3481 \eta}{168} + 43 \eta^2\right) \Sigma_{\ell}^2\right) v^2 \\
    & \quad + \left(32 \pi S_{\ell}^2 + 32 \delta \pi S_{\ell} \Sigma_{\ell} + \left(\frac{65 \pi}{8} - 32 \eta \pi\right) \Sigma_{\ell}^2\right) v^3 ~,
    \end{aligned}
\end{equation}
while the spin-cubic interaction is given by
\begin{equation}
    \mathcal{F}_{\rm SSS}(v) = -\frac{16}{3} S_{\ell}^3 + \frac{2}{3} \delta S_{\ell}^2 \Sigma_{\ell} + \left(\frac{9}{2} - \frac{56 \eta}{3}\right) S_{\ell} \Sigma_{\ell}^2 + \left(\frac{35 \delta}{24} - 6 \delta \eta\right) \Sigma_{\ell}^3 ~.
\end{equation}

\vskip 2pt

\subsection*{Energy Flux from Tidal Heating}

From the effective action \eqref{eq:multi_tides}, we derive the tidal heating energy flux of a single body, which is equivalent to the equation of motion of the effective action. For a general tidal environment, this energy flux is
\begin{equation}
    \begin{aligned}
    \dot{m} & =  - m(G m)^4 \Bigg[ H_{S}^E \chi \( \dot E^{ij} E_{i}{}^k \hat{S}_{jk}\) - (Gm) H_{ \omega}^E \dot{E}^{ij} \dot{E}_{ij} \\
    & \quad +  H_{S}^B \chi \( \dot B^{ij} B_{i}{}^k \hat{S}_{jk}\) - (Gm) H_{ \omega}^B \dot{B}^{ij} \dot{B}_{ij} \Bigg] ~.
    \end{aligned} \label{eqn: mdot_single}
\end{equation}
and angular momentum flux
\begin{equation}
    \begin{aligned}
        \dot J & =- m(Gm)^4 \Bigg[ -2 H_{S}^E \chi \left(E^{ij} E_{ij}\right) + 3 H_{S}^E \chi  \left(E_i{ }^k E_{j k} \hat{s}^i \hat{s}^j \right) + (GM) H_{\omega}^{E} (\dot E^{ij} E_{i}{}^k \hat{S}_{j k}) \\
& \quad -2 H_{S}^B \chi \left(B^{ij} B_{ij} \right) + 3 H_{S}^B \chi  \left(B_i{ }^k B_{j k} \hat{s}^i \hat{s}^j\right) + (GM) H_{\omega}^{B} (\dot{B}^{ij} B_{i}{}^k \hat{S}_{j k}) \Bigg] ~,
    \end{aligned}
\end{equation}
where the $J$ is the dimensionful spin $J\equiv G m^2 \chi$.
For a quasi-circular aligned-spin orbit, the tidal moments $E_{ij}$ and $B_{ij}$ have been computed up to $\mathcal{O}(v^3)$ and $\mathcal{O}(v)$ beyond leading order~\cite{Taylor:2008xy, Chatziioannou:2012gq, Chatziioannou:2016kem}, which are sufficiently high PN order for our purposes; see (30)-(34) in~\cite{Chatziioannou:2016kem}. Substituting these tidal moments into (\ref{eqn: mdot_single}), we obtain the tidal heating flux for $m_1$:
\begin{equation}
\begin{aligned}
\dot{m_1}(v) & = \left( \frac{{9 H_{1S}^{E} m_1^3 \eta^2 \chi_1}}{{M^3}} \right) v^{15} + \Bigg( -\frac{{9 H_{1S}^{E} m_1^3 (-2 M - m_1 + 3 M \eta) \eta^2 \chi_1}}{{M^4}} \\
& + \frac{{9 H_{1S}^{B} m_1^3 \eta^2 \chi_1}}{{M^3}} \Bigg) v^{17} + \left( \frac{{18 H_{1\omega} m_1^4 \eta^2}}{{M^4}} \right) v^{18} ~.
\end{aligned} \label{eqn: mdot_single_circle}
\end{equation}
The expression for $\dot{m_2}$ is obtained by interchanging the $1 \leftrightarrow 2$ indices. The total tidal heating flux absorbed by both binary components, $\dot{M} = \dot{m_1} + \dot{m_2}$, is therefore 
\begin{equation}
    \begin{aligned}
        \dot M (v) & = \left( 9 \overline{\mathcal{H}}_1^E \eta^2 \chi_a + 9 \mathcal{H}_1^E \eta^2 \chi_s \right) v^{15} + \Bigg[\Bigg(9 \mathcal{H}_1^B \eta^2 + \frac{45 \mathcal{H}_1^E \eta^2}{2} + \frac{9}{2} \overline{\mathcal{H}}_1^E \delta \eta^2 - 27 \mathcal{H}_1^E \eta^3\Big) \chi_s \\
        & \quad + \Bigg(9 \overline{\mathcal{H}}_1^B \eta^2 + \frac{45 \overline{\mathcal{H}}_1^E \eta^2}{2} + \frac{9}{2} \mathcal{H}_1^E \delta \eta^2 - 27 \overline{\mathcal{H}}_1^E \eta^3 \Bigg) \chi_a \Bigg] v^{17} + 18 \mathcal{H}_0 \eta^2 v^{18}  ~,  \label{eqn: mdot_binary_circle}
    \end{aligned}
\end{equation}
where $\mathcal{H}^{E/B}_{1}, \overline{\mathcal{H}}^{E/B}_{1}$ and $\mathcal{H}^{E/B}_{0}$ are the effective dissipation numbers defined in (\ref{eqn:all_effective_dissipation_numbers}).
When studying binary black holes, (\ref{eqn: mdot_single_circle}) and (\ref{eqn: mdot_binary_circle}) can be simplified via the electric-magnetic duality, $\mathcal{H}^{E}_{1}=\mathcal{H}^{B}_{1}$, $\overline{\mathcal{H}}^{E}_{1}=\overline{\mathcal{H}}^{B}_{1}$ and $\mathcal{H}^{E}_{0}=\mathcal{H}^{B}_{0}$ described in Section \ref{sec:theory}, with their theoretical values found in \eqref{eqn:bh_dissipation_general}.

\subsection*{GW Phase}

Using the results shown above, we may derive the analytic expressions for waveform observables. In particular,  the derivation for both the phase and amplitude utilize the energy balance equation:
\begin{equation}
   - \mathcal{F}_\infty - \dot M = \dot{\mathcal{E}}  \, . \label{eqn:Ebalance}
\end{equation}
To compute the phase, one has to solve for the GW orbital phase $\phi(v)$ and time $t(v)$ as observed at asymptotic infinity~\cite{Buonanno:2009zt}. Since $\tau \to t$ at asymptotic infinity, in what follows we reuse overdot for derivative in $t$ for notational simplicity. For quasi-circular orbits, the orbital phase can be solved via Kepler's third law, which gives $\dot{\phi} = v^3 / M$ for the $m_\phi = 2$ radiation mode. Integrating over the equations interatively, we have
\begin{equation}
\begin{aligned}
t(v)  = t_0 + \int \d v  \, \frac{ 1}{\dot{v}} \, , 
\qquad \phi(v) = \phi_0 + \int \d v \, \frac{v^3}{M} \frac{1}{\dot{v}} \, . \label{eqn:phase_time}
\end{aligned}
\end{equation}
The expression for $\dot{v}$ can be obtained by applying chain rule to $\dot{\mathcal{E}}$ in (\ref{eqn:Ebalance}) -- see also (\ref{eqn:Ebalance_chainrule_maintext}) -- and reorganizing it as follows:
\begin{equation}
 \dot{v} = - \left( \frac{\partial \mathcal{E}}{\partial v} \right)^{-1} \left[ \mathcal{F}_\infty + \left( 1 - \frac{\partial \mathcal{E}}{\partial m_1} \right) \dot{m_1} + \left( 1 - \frac{\partial \mathcal{E}}{\partial m_2} \dot{m_2} \right) \right] ~ , \label{eqn:Ebalance_chainrule}
\end{equation}
whose explicit result for quasicircular orbits can be computed by substituting (\ref{eqn:binding}), (\ref{eqn:flux}) and (\ref{eqn: mdot_single_circle}). As described in the main text, in addition to the fluxes $\dot{m_1}$ and $\dot{m_2}$ which are directly absorbed by the binary components, there are also the so-called ``secular change in mass" terms arising from applying chain rule to the energy balance equation. Since the leading binding energy scales as $\mathcal{E} \propto v^2$, cf. (\ref{eqn:binding}), these secular corrections are 1PN suppressed compared $\dot{m}_{1,2}$.

\vskip 4pt

Solving (\ref{eqn:phase_time}) iteratively after performing a Taylor expansion on (\ref{eqn:Ebalance_chainrule}), we arrive at the TaylorF2 waveform phase for the $(2,2)$ radiation mode, where $\psi(v) = 2 \pi f t(v) - 2 \phi(v)$ is the phase in the stationary phase approximation and $v = (\pi M f)^{1/3}$cf. (\ref{eqn:freq_amplitude}). We separate the phase into the point particle (black hole) and the tidal dissipation number contributions:
\begin{equation}
    \psi(v) = 2\pi f t_0 - \phi_0 - \frac{\pi}{4} + \frac{3}{128\eta v^5} \Big[  \psi^{\rm pp}(v) + \psi^{\rm TDN} (v)  \Big] ~,
\end{equation}
where the tidal dissipation terms are shown in (\ref{eq: dissipation phase}) and (\ref{eqn:dissipation_phase_PN}) , while the point particle phase contribution is
\begin{equation}
    \begin{aligned}
    \psi^{\rm pp}(v) & = 1 + v^2 \psi_{\rm 1PN}^{\rm pp} + v^3 \psi_{\rm 1.5PN}^{\rm pp} + v^4 \psi_{\rm 2PN}^{\rm pp} + v^5 (1 + 3 \ln v)\psi_{\rm 2.5PN}^{\rm pp} + v^6 \psi_{\rm 3PN}^{{\rm pp}}  \\
    & \quad + v^6 \ln v \hskip 2pt \psi_{\rm 3PN}^{{\rm pp, ln} }  + v^7 \psi_{\rm 3.5 PN}^{\rm pp} + v^{8}(1-3\ln v) \psi_{\rm 4PN}^{\rm pp} + v^8 \left( \ln v\right)^2  \hskip 2pt \psi_{\rm 4PN}^{\rm pp, ln^2} ~,  \label{eqn:phase_PP_general}
    \end{aligned}
\end{equation}
with the PN coefficients~\cite{Blanchet:2023bwj, Cho:2022syn}
\begin{align*}
     \psi_{\rm 1PN}^{\rm pp} & = \frac{3715}{756}+\frac{55 \eta}{9}, \\
     \psi_{\rm 1.5PN}^{\rm pp} & = -16 \pi+\left(\frac{113}{3}-\frac{76 \eta}{3}\right) \chi_s + \frac{113 \delta \chi_a}{3} ~, \\
     \psi_{\rm 2PN}^{\rm pp} & = \frac{15293365}{508032} + \frac{27145 \eta}{504} + \frac{3085 \eta^2}{72} - \left(\frac{405}{8} - 200 \eta\right) \chi_a^2 - \frac{405 \delta \chi_a \chi_s}{4} - \left(\frac{405}{8} - \frac{5 \eta}{2}\right) \chi_s^2 ~, \\
     \psi_{\rm 2.5PN}^{\rm pp}  & = \frac{38645 \pi}{756}-\frac{65 \pi \eta}{9} + \( - \frac{732985}{2268} + \frac{24260}{81}\eta + \frac{340}{9} \eta^2 \) \chi_s + \( - \frac{732985}{2268}\delta - \frac{140}{9}\delta \eta \) \chi_a ~, \\
    \psi_{\rm 3PN}^{{\rm pp}} & = \frac{11583231236531}{4694215680} - \frac{6848 \gamma_E}{21} - \frac{640 \pi^2}{3} - \left(\frac{15737765635}{3048192} - \frac{2255 \pi^2}{12}\right) \eta + \frac{76055 \eta^2}{1728} \\
        & \quad - \frac{127825 \eta^3}{1296} + \frac{2270 \pi \delta \chi_a}{3} + \left(\frac{75515}{288} - \frac{263245 \eta}{252} - 480 \eta^2\right) \chi_a^2 \\
        & \quad + \left(\frac{2270 \pi}{3} - 520 \pi \eta + \left(\frac{75515 \delta}{144} - \frac{8225 \delta \eta}{18}\right) \chi_a\right) \chi_s \\
        & \quad + \left(\frac{75515}{288} - \frac{232415 \eta}{504} + \frac{1255 \eta^2}{9}\right) \chi_s^2 - \frac{13696 \ln 2}{21} ~, \tag{\stepcounter{equation}\theequation} \\
        \psi_{\rm 3PN}^{{\rm pp, ln}} & = - \frac{6848}{21} ~,  \\
     \psi_{\rm 3.5PN}^{\rm pp} & = \frac{77096675 \pi}{254016} + \frac{378515 \pi \eta}{1512} - \frac{74045 \pi \eta^2}{756} + \left(-\frac{25150083775 \delta}{3048192} + \frac{26804935 \delta \eta}{6048} - \frac{1985 \delta \eta^2}{48}\right) \chi_a \\
    & \quad + \left(-\frac{815 \pi}{2} + 1600 \pi \eta\right) \chi_a^2 + \left(\frac{14585 \delta}{24} - 2380 \delta \eta\right) \chi_a^3 \\ 
    & \quad + \Bigg(-\frac{25150083775}{3048192} + \frac{10566655595 \eta}{762048} - \frac{1042165 \eta^2}{3024} + \frac{5345 \eta^3}{36} \\
    & \quad - 815 \pi \delta \chi_a + \left(\frac{14585}{8} - 7270 \eta + 80 \eta^2\right) \chi_a^2\Bigg) \chi_s \\
    & \quad + \left(-\frac{815 \pi}{2} + 30 \pi \eta + \left(\frac{14585 \delta}{8} - \frac{215 \delta \eta}{2}\right) \chi_a\right) \chi_s^2 + \left(\frac{14585}{24} - \frac{475 \eta}{6} + \frac{100 \eta^2}{3}\right) \chi_s^3 ~,  \\
    \psi_{\rm 4PN}^{\rm pp} & = -\frac{90490\pi ^2}{567}-\frac{36812 \gamma_E }{189}+\frac{2550713843998885153}{830425530654720}-\frac{26325 \ln 3}{196}-\frac{1011020 \ln 2}{3969} \\
& \quad +\eta  \left(-\frac{680712846248317}{126743823360}-\frac{3911888 \gamma_E }{3969}+\frac{109295 \pi ^2}{672}-\frac{9964112 \ln 2}{3969}+\frac{26325 \ln 3}{49}\right) \\
& \quad + \left(\frac{7510073635}{9144576}-\frac{11275 \pi ^2}{432}\right) \eta ^2 +\frac{1292395 \eta ^3}{36288} -\frac{5975 \eta ^4}{288} ~, \\
\psi_{\rm 4PN}^{\rm pp, ln^2} & = \frac{1955944 \eta }{1323}+\frac{18406}{63} ~.
\end{align*}
Note that the constant $``1"$ terms in the $(1 \pm 3 \ln v)$ factors in (\ref{eq: dissipation phase}) and (\ref{eqn:phase_PP_general}) at 2.5PN and 4PN orders are occasionally ignored in the literature as they can be absorbed into the redefinition of $\phi_0$ and $t_0$ respectively. On the other hand, the $\pm 3\ln v$ terms cannot be absorbed into redefinitions and arise from integrations of the form $\int dv/v$ when computing $\phi(v)$ and $t(v)$ in (\ref{eqn:phase_time}). The presence of these logarithms is the reason why the 2.5PN and 4PN coefficients are still important in the PN waveform model.

\vskip 2pt

\subsection*{Energy Loss}
With the flux radiated to infinity given in Eq.~\eqref{eqn:flux} and the flux at the horizon in Eq.~\eqref{eqn: mdot_binary_circle}, we may also derive the analytic expression for the energy loss during the inspiral phase:
\begin{equation}
    \Delta E_{\infty} (v)= \int \mathcal{F}_{\infty} \frac{dv}{\dot v} ~, \quad \Delta E_{\rm H} (v) = \int \dot M \frac{dv}{\dot v} ~. \label{eqn:energy_loss}
\end{equation}
In the above expression, $\dot v$ is given in Eq.~\eqref{eqn:Ebalance_chainrule}. Solving this iteratively, we get the energy loss at infinity
\begin{equation}
\begin{aligned}
    \Delta E_{\infty}(v) & = \frac{1}{2} M v^2 \eta \Big[1 + \Delta E_{\infty}^{\rm 1PN} v^2 + \Delta E_{\infty}^{\rm 1.5PN} v^3 + \Delta E_{\infty}^{\rm 2PN} v^4 + \Delta E_{\infty}^{\rm 2.5PN} v^5 \\
    & \quad + \Delta E_{\infty}^{\rm 3PN} v^6 + \Delta E_{\infty}^{\rm 3.5PN} v^7 + \Delta E^{\rm 4PN}_\infty v^8 + \Delta E^{\rm 4PN,\ln}_{\infty} v^8 \ln v \Big] ~,
\end{aligned}
\end{equation}
where
\begin{equation}
\begin{aligned}
\label{eq: energy loss inf}
     \Delta E_{\infty}^{\rm 1PN} & = - \frac{3}{4} - \frac{\eta}{12} ~, \quad \Delta E_\infty^{\rm 1.5PN} = \frac{8}{3} \delta \chi_a + \frac{8}{3} \chi_s - \frac{4}{3} \eta \chi_s ~, \\
     \Delta E_{\infty}^{\rm 2PN} & = - \frac{27}{8} + \frac{19}{8} \eta - \frac{\eta^2}{24} + (-1 + 4 \eta) \chi_a^2 - 2 \delta \chi_a \chi_s - \chi_s^2 ~, \\
     \Delta E_{\infty}^{\rm 2.5PN} & = \Bigg(8 \delta - \frac{31}{9} \delta \eta\Bigg) \chi_a + \(8 - \frac{121}{9} \eta + \frac{2}{9}\eta^2\) \chi_s - \frac{45}{112} \bar{\mathcal{H}}_1^E \chi_a - \frac{45}{112} \mathcal{H}_1^E \chi_s ~, \\
     \Delta E_{\infty}^{\rm 3PN} & =  - \frac{675}{64} + \(\frac{34445}{576} - \frac{205}{96} \pi^2\) \eta - \frac{155}{96} \eta^2 - \frac{35}{5184} \eta^3 \\
    & \quad + \(- \frac{65}{18} + \frac{305}{18}\eta - \frac{10}{3} \eta^2\)\chi_a^2 + \(-\frac{65}{9} \delta + \frac{145}{9} \delta \eta\) \chi_a \chi_s + \(-\frac{65}{18} + \frac{245}{18} \eta - \frac{40}{9} \eta^2\) \chi_s^2 ~,  \\
     \Delta E_\infty^{\rm 3.5PN} & = \(27 \delta - \frac{211}{4} \delta \eta + \frac{7}{6} \delta \eta^2 \)\chi_a + \(27 - \frac{373}{4} \eta + \frac{86}{3} \eta^2 + \frac{\eta^3}{6}\) + (-4 \eta + 16 \eta^2)\chi_a^2 \chi_s \\
    & \quad - 8 \delta \eta \chi_a \chi_s^2 - 4 \eta \chi_s^3 + \( \(-\frac{7495}{5376} - \frac{5}{64} \eta\) \chi_a - \frac{15}{64} \delta \chi_s\) \bar{\mathcal{H}}_1^E- \frac{5}{16} \chi_a \bar{\mathcal{H}}_1^B \\
    & \quad + \(\(-\frac{7495}{5376} - \frac{5}{64} \eta\) \chi_s - \frac{15}{64} \delta \chi_a\) \mathcal{H}_1^E - \frac{5}{16} \chi_s \mathcal{H}_1^B ~, \\
    \quad \Delta E_\infty^{\rm 4PN} & = - \frac{3969}{128} + \eta \( - \frac{123671}{5760} + \frac{896}{15} \gamma_E + \frac{9037}{1536}\pi^2 + \frac{1792}{15} \log(2)\) + \( - \frac{498449}{3456} + \frac{3157}{576} \pi^2\) \eta^2 \\
    & \quad + \frac{301}{1728} \eta^3 + \frac{77}{31104} \eta^4 - \frac{9}{16} \mathcal{H}_0 ~, \\
    \Delta E_\infty^{\rm 4PN,\ln} & = \frac{896}{15} ~.
\end{aligned}
\end{equation}
The energy loss at the horizon is given by
\begin{equation}
\label{eq: energy loss horizon}
    \Delta E_{\rm H}(v) = \frac{1}{2} M v^2 \eta \Big[ \Delta E_{\rm H}^{\rm 2.5PN} v^5 + \Delta E_{\rm H}^{\rm 3.5PN} v^7 + \Delta E_{\rm H}^{\rm 4PN} v^8 \Big]
\end{equation}
where
\begin{equation}
\begin{aligned}
    \Delta E_{\rm H}^{\rm 2.5PN} & = \frac{45}{112} \mathcal{H}_1^E \chi_s + \frac{45}{112} \bar{\mathcal{H}}_1^E \chi_a ~, \\
    \Delta E_{\rm H}^{\rm 3.5PN} & = \(\frac{5}{16} \mathcal{H}_1^B + \frac{7915}{5376} \mathcal{H}_1^E + \frac{5}{32} \bar{\mathcal{H}}_1^E \delta - \frac{5}{64} \mathcal{H}_1^E \eta \) \chi_s \\
    & \quad + \( \frac{5}{16} \bar{\mathcal{H}}_1^B + \frac{7915}{5376} \bar{\mathcal{H}}_1^E + \frac{5}{32} \mathcal{H}_1^E \delta - \frac{5}{64} \bar{\mathcal{H}}_1^E \eta \) \chi_a ~, \\
    \Delta E_{\rm H}^{\rm 4PN} & = \frac{9}{16} \mathcal{H}_0 ~.
\end{aligned}
\end{equation}

\vskip 2pt

\subsection*{GW Amplitude}

In the stationary phase approximation~\cite{Thorne1980Lectures}, the Fourier domain waveform amplitude defined in (\ref{eqn:freq_amplitude}) takes the form~\cite{VanDenBroeck:2006qu, Buonanno:2009zt, Maggiore:2007ulw}
\begin{equation}
A(v) = \frac{2 \eta M}{D} v^2 \left( \dot{F} (v) \right)_{F=f}^{-1/2} \, , 
\end{equation}
where $( \dot{F}(v))_{F=f}$ is the time derivative of the instantaneous GW frequency evaluated at the stationary point, $F = f = v^3 / (\pi M)$. Using chain rule and the energy balance equation (\ref{eqn:Ebalance}), one can re-express the instantaneous frequency derivative and the amplitude as
\begin{equation}
   \dot{F}(v) = \frac{\partial F}{\partial v} \frac{\partial v}{\partial \mathcal{E}} \dot{\mathcal{E}} \, , \qquad A(v)  = \frac{2\eta M^{3/2} v}{D} \sqrt{\frac{\pi}{3}} \[ \frac{- \partial \mathcal{E} / \partial v }{\mathcal{F}_{\infty} + \dot M}\]^{1/2} \, , 
\end{equation}
Subsituting (\ref{eqn:binding}), (\ref{eqn:flux}) and (\ref{eqn: mdot_binary_circle}) for the binding energy and fluxes, one obtains
\begin{equation}
\begin{aligned}
    A(f) & = \sqrt{\frac{5}{24}} \frac{\mathcal{M}^{5 / 6}}{D\pi^{2/3}} f^{-7 / 6} \Big[ A^{\rm pp}(f) + A^{\rm TDN}(f) \Big] ~,
\end{aligned}
\end{equation}
where the amplitude corrections from tidal dissipation effects are shown in (\ref{eqn:dissipation_amplitudes}) and (\ref{eqn:dissipation_amplitudes_PN}), while the point particle contributions are 
\begin{equation}
    \begin{aligned}
        A^{\rm pp}(f) & = 1 + v^2 A_{\rm 1PN}^{\rm pp} + v^3 A_{\rm 1.5PN}^{\rm pp} + v^4 A_{\rm 2PN}^{\rm pp} + v^{5}A_{\rm 2.5PN}^{\rm pp} + v^6 A_{\rm 3PN}^{\rm pp} \\
        & \quad  + v^6 \ln v A_{\rm 3PN}^{\rm pp, ln} + v^7 A_{\rm 3.5PN}^{\rm pp} + v^8 A_{\rm 4PN}^{\rm pp} + v^8 \ln v  A_{\rm 4PN}^{\rm pp, ln}  ~,  \label{eqn:amplitude_PP_general}
    \end{aligned}
\end{equation}
with the PN coefficients
\begin{align*}
A_{\rm 1PN}^{\rm pp} & = \frac{11}{8} \eta + \frac{743}{672} \\
A_{\rm 1.5PN}^{\rm pp} & = -2\pi + \frac{113}{24} \delta \chi_a + \frac{113}{24} \chi_s - \frac{19}{6} \eta \chi_s \\
A_{\rm 2PN}^{\rm pp} & = \frac{7266251}{8128512} + \frac{18913 \eta}{16128} + \frac{1379 \eta^2}{1152} - \left(\frac{81}{32} - 10 \eta\right) \chi_a^2 - \frac{81 \delta \chi_a \chi_s}{16} - \left(\frac{81}{32} - \frac{\eta}{8}\right) \chi_s^2 ~, \\
 A_{\rm 2.5PN}^{\rm pp} & = -\frac{4757 \pi}{1344} + \frac{57 \pi \eta}{16} + \left(\frac{502429 \delta}{16128} - \frac{907 \delta \eta}{192}\right) \chi_a + \left(\frac{502429}{16128} - \frac{73921 \eta}{2016} + \frac{5 \eta^2}{48}\right) \chi_s ~, \\
 A_{\rm 3PN}^{\rm pp} & = -\frac{29342493702821}{500716339200} + \frac{856 \gamma_E}{105} + \frac{10 \pi^2}{3}  + \frac{1712 \ln 2}{105}  + \left(\frac{3526813753}{27869184} - \frac{451 \pi^2}{96}\right) \eta  \\
    & \quad - \frac{1041557 \eta^2}{258048} + \frac{67999 \eta^3}{82944} - \frac{19 \pi \delta \chi_a}{2} + \left(-\frac{19 \pi}{2} + \frac{20 \pi \eta}{3} \right)\chi_s \\
    & \quad + \left(-\frac{319133}{21504} + \frac{48289 \eta}{768} - \frac{7 \eta^2}{4}\right) \chi_a^2 + \left(- \frac{319133 \delta}{10752} + \frac{12785 \delta \eta}{384}  \right) \chi_a \chi_s \\
    & \quad   + \left(-\frac{319133}{21504} + \frac{40025 \eta}{1344} - \frac{555 \eta^2}{64}\right) \chi_s^2 ~ , \tag{\stepcounter{equation}\theequation}  \\
A_{\rm 3PN}^{\rm pp, ln} & = \frac{856}{105} ~,   \\
A_{\rm 3.5PN}^{\rm pp} & = -\frac{5111593 \pi}{2709504} - \frac{72221 \pi \eta}{24192} - \frac{1349 \pi \eta^2}{24192} + \left(\frac{1557122011 \delta}{9289728} - \frac{2206797 \delta \eta}{14336} + \frac{52343 \delta \eta^2}{27648}\right) \chi_a \\
    & \quad + \left(\frac{41 \pi}{8} - 20 \pi \eta\right) \chi_a^2 + \left(-\frac{2515 \delta}{768} + \frac{149 \delta \eta}{12}\right) \chi_a^3 + \Bigg[\frac{1557122011}{9289728} - \frac{1905526039 \eta}{5419008} \\
    & \quad + \frac{11030651 \eta^2}{193536} - \frac{445 \eta^3}{6912} + \frac{41 \pi \delta \chi_a}{4} + \left(-\frac{2515}{256} + \frac{6011 \eta}{192} + \frac{89 \eta^2}{3}\right) \chi_a^2\Bigg] \chi_s \\
    & \quad + \left(\frac{41 \pi}{8} - \frac{\pi \eta}{2} + \left(-\frac{2515 \delta}{256} - \frac{2675 \delta \eta}{192}\right) \chi_a\right) \chi_s^2 - \left(\frac{2515}{768} + \frac{53 \eta}{8} + \frac{7 \eta^2}{16}\right) \chi_s^3 ~, \\
A_{\rm 4PN}^{\rm pp} & = -\frac{246427872050556151}{899847347503104} + \frac{56881 \gamma_E}{4410} + \frac{14495 \pi^2}{2016} + \left(-\frac{2469217875055}{9364045824} + \frac{451 \pi^2}{48}\right) \eta^2 \\
    & \quad - \frac{42749765 \eta^3}{55738368} + \frac{144587 \eta^4}{294912} + \frac{10417 \ln 2}{980} + \frac{47385 \ln 3}{3136}  \\
    & \quad + \eta \left(\frac{997052025430343}{1716741734400} + \frac{219776 \gamma_E}{2205} - \frac{681325 \pi^2}{64512} + \frac{573323 \ln 2}{2205} - \frac{47385 \ln 3}{784}\right) ~, \\
A_{\rm 4PN}^{\rm pp,ln } & = \( \frac{56881}{4410} + \frac{219776}{2205}\eta\) ~.
\end{align*}

\phantomsection
\addcontentsline{toc}{section}{References}
\bibliographystyle{utphys}
\bibliography{references.bib}
\end{document}